\documentclass[12pt]{article}
\usepackage{latexsym}
\usepackage{epsfig}

\usepackage{graphicx}
\usepackage{epstopdf}

\usepackage{epsfig,amssymb,amsmath,amscd}

\newcommand{\bq}{\begin{equation}}
\newcommand{\eq}{\end{equation}}
\newcommand{\bqa}{\begin{eqnarray}}
\newcommand{\eqa}{\end{eqnarray}}
\newcommand{\ben}{\begin{enumerate}}
\newcommand{\een}{\end{enumerate}}
\newcommand{\bc}{\begin{center}}
\newcommand{\ec}{\end{center}}
\newcommand{\bqb}{\begin{eqnarray*}}
\newcommand{\eqb}{\end{eqnarray*}}

\def\pr#1#2#3{Phys. Rev. ${\bf{#1}}$, #2 (#3)}

\def\pl#1#2#3{Phys. Lett. ${\bf{#1}}$, #2 (#3)}

\def\np#1#2#3{Nucl. Phys. ${\bf{#1}}$, #2 (#3)}

\def\jhep#1#2#3{JHEP ${\bf{#1}}$, #2 (#3)}
\def\epj#1#2#3{Eur. Phys. J. ${\bf{#1}}$, #2 (#3)}

\def\jmp#1#2#3{J. Mod. Phys. ${\bf{#1}}$, #2 (#3)}

\begin{document}
\pagenumbering{arabic}
\thispagestyle{empty}
\def\thefootnote{\fnsymbol{footnote}}
\setcounter{footnote}{1}

\begin{flushright}
Sept 10, 2017\\
 \end{flushright}

\begin{center}
{\Large {\bf CSM tests in $gb\to \gamma b, Hb, Zb, W^-t$ processes}}.\\
 \vspace{1cm}
{\large F.M. Renard}\\
\vspace{0.2cm}
Laboratoire Univers et Particules de Montpellier,
UMR 5299\\
Universit\'{e} Montpellier II, Place Eug\`{e}ne Bataillon CC072\\
 F-34095 Montpellier Cedex 5, France.\\
\end{center}

\vspace*{1.cm}
\begin{center}
{\bf Abstract}
\end{center}

We show that the processes of $\gamma b, Hb, Zb, W^-t$ production
in gluon+bottom collision can give interesting informations about possible
Higgs boson, top and bottom quark compositeness. We make illustrations
of the ratios of new cross sections over standard ones. Specific
energy dependences appear for each assumption about $b_L$, $b_R$,
$t_L$, $t_R$ compositeness and CSM constraints concerning the
Higgs sector.

\vspace{0.5cm}

\def\thefootnote{\arabic{footnote}}
\setcounter{footnote}{0}
\clearpage

\section{INTRODUCTION}

We have recently looked at the effects of
Higgs boson, top and bottom quark compositeness in several
processes occuring in $e^+e-$, photon-photon and hadronic collisions.
The motivation was essentially to test the concept of
Compositeness Standard Model (CSM), see ref.\cite{CSMrev,CSMbot}.\\

This concept consists in assuming that the SM can be constructed
in a simple way, for example starting from substructures like in \cite{comp},
and that its main properties are preserved at low energies. 
The first CSM effects could be the appearence of form factors 
(but no anomalous coupling) and of effective s-dependent masses.
The reproduction of the SM structures implies the preservation of the 
Goldstone equivalence with the longitudinal gauge boson amplitudes
(we will often denote this property as CSMG).\\

General compositeness of the top quark and of the Higgs boson 
has been studied in \cite{partialcomp,Hcomp2,Hcomp3,Hcomp4}.
The observability of top compositeness has also been discussed in
\cite{Tait}.
In our studies we wanted to see how one could immediately differentiate  
compositeness effects corresponding to CSM conservation from those 
corresponding to
CSM violation.
A strategy starting from the detection of form factor effects
in simple processes and then pursuing with more involved processes
producing gauge and Higgs bosons, top and bottom quarks, 
has been proposed in \cite{CSMrev,CSMbot}.\\

In this short paper we just want to add a few more tests
realizable with the $gb\to \gamma b, Hb, Zb, W^-t$ processes.
This is done in Section 2 where we give detailed illustrations for each process
and each compositeness assumption. A summary is given in Section 3.\\

\section{$gb \to Xf$ Processes}

We consider the four processes corresponding to $Xf \equiv \gamma b, Hb, Zb, W^-t$.
At Born level they occur through the s-channel and u-channel diagrams
drawn in Fig.1.\\
With compositeness the point-like couplings may be replaced
by effective s-dependent quantities that we represent by 
test form factors of the type:

\bq
F(s)={s_0+M^2\over s+M^2}~~\label{FF}
\eq
\noindent
with the new physics scale $M$ taken for example in the few TeV range.\\
We will compute the effects of such form factors on the cross sections of the
various processes and show them by drawing the ratios of the new cross 
sections over the SM ones. 
We will illustrate their energy dependence (for an arbitrary scattering 
angle of 30 degrees) and when they are important, their angular
dependence which is essentially generated by the u-channel exchange term.\\
In these illustrations we will use the following notations,
tL, tR, tLR for pure $t_L$ or $t_R$ or both form factors; tLm, tRm, tLRm for form factors
together with effective $m_t(s)$ mass; and similarly bL, bR, bLR, bLm, bRm, BLRm
for bottom compositeness and tLbL, tRbR, tLRbLR, tLmbLm, tRMbRm, tLRmbLRm 
for both top and  bottom compositeness.\\
As discussed for example in ref.\cite{Hcomp4} there is the
possibility of mixing of elementary states with composite ones.
We will illustrate the full compositeness cases. 
Partial compositeness should give intermediate effects obtained by
factorizing the mixing angles.\\

\underline{$gb\to \gamma b$}\\

After $e^+e^- \to b \bar b$ this process may be also interesting for providing
direct simple tests of bottom compositeness.
It will allow to test the presence of $gbb$ and $\gamma bb$ form
factors. 
If these form factors arise from bottom substructure they could
depend on the colour and on the electric charge of the constituents, such that
$gbb$ and $\gamma bb$ form factors may be different.\\
In Fig.2 we illustrate the effects of the choice of eq.(\ref{FF}) for pure 
$b_L$ compositeness or pure $b_R$ compositeness or for both. For simplicity
we take the same form factors for $gbb$ and $\gamma bb$ couplings.
The angular distribution of the ratios is constant in this case. \\
With only transverse real photons there is no visible bottom mass effect.\\

\underline{$gb\to Hb$}\\

This process makes one more step as it involves in addition to the  $gbb$ form
factor, a sensitivity to the bottom mass appearing in the $Hbb$
coupling.\\
As introduced in \cite{trcomp} compositeness may generate an effective 
s-dependent bottom mass.\\
This $Hbb$ coupling appears in the left and in the right terms of
the s- and u- channel diagrams  which combine and partially cancel in
the SM case. 
So when one introduces different $b_L$ or $b_R$
modifications this affects the cancellations and leads to an increase
of the cross section as one can see with the ratios drawn in Fig.3(up).
These effects are strongly angular dependent essentially due to the 
u-channel contributions
and if a deviation from SM is observed the study of its angular distribution
should be instructive; see Fig.3(down).\\
We can observe the separate effects of gluon form factors for $b_L$,
for $b_R$ or for both and similarly the additional effect of an effective mass
$m_b(s)=m_bF(s)$.\\  
One can also check that the SM cancellations are recovered
when both $b_L$ and $b_R$ are affected by the same form factor such that, 
in this case, the ratios decrease strongly with the energy.\\

\underline{$gb\to Z b$}\\

We first treat separately the transverse $Z_T$ and the longitudinal $Z_L$
production 	as illustrated in Fig.4.
A priori the $Z_T b$ case should be rather similar to the above $ \gamma b$
one. This is true apart from the fact that the $Z_Rbb$ coupling is smaller
than the $Z_Lbb$ one (whereas they were equal in the photon case)
such that the $b_L$ and $b_R$ curves now differ, see Fig.4(up).\\
The $Z_L$ case is however much more informative. There appears now a
big sensitivity to the bottom mass which arises after the typical SM
cancellation of the longitudinal amplitudes leading to an $m_b$ term 
in agreement with the Goldstone equivalence which predicts, up to
$m^2_Z/s$ terms, that the $gb\to Z_L b$ amplitude should be equal to
the $gb\to G^0 b$ one. In Fig.4(middle) one can see the 
separate sensitivity of the cross section ratios to the $b_{L,R}$ 
form factors and to the $m_b(s)$ effective mass.\\
In Fig.4(down) we show what would be the influence of the form factors
and of the effective bottom mass on $Z_L$ production if the substructure 
effects respect the Goldstone equivalence as required by the CSMG
assumption.\\
All these informations would be very interesting, however the rate
of $Z_L$ production controlled by $m_b$
(less than 1 percent of the total $Z$ rate at low energy and decreasing
strongly with the energy) will probably
not allow their observability. Only the unpolarized case, with effects
similar to the $Z_T$ ones shown in Fig.4(up) may be observable.\\
Hopefully the $W$ production process, that we will now study, should be more adequate in
this respect due to the larger top mass.\\

\underline{$gb\to W^- t$}\\

For this process we will make 3 types of studies corresponding to 
the effects of top or of bottom compositeness or of both.
In each case we will also separate the effect of pure Left compositeness, 
of pure Right compositeness and of both.\\   
We will look at the effects on $W^-_T$,  on $W^-_L$ and on the unpolarized 
$W^-$ production.
As expected the $W^-_T$ ratios are not sensitive to the top and bottom
effective masses and allow to only test the presence of the form
factors in the couplings, essentially the left-handed ones which
appear with the pure Left W couplings. On the opposite the $W^-_L$ ratios
are very sensitive to the effective masses (essentially the top one)
because they control
the resulting quantities after the cancellation of the usual increasing
(unitarity violating) contributions to the longitudinal amplitudes.
Because of these properties the $W^-_L$ contributions are now important and
lead also to modifications of the unpolarized $W^-$ cross sections as we can see 
in the following figures.\\

\underline{Effects of pure t compositeness}\\

In Fig.5(up, middle, down) one sees the effects of $t_{L,R}$ compositeness
on the $W^-_T$, $W^-_L$ and $G^-$ ratios. One can also see the effects of
an effective s-dependent top mass on the $W^-_L$ and $G^-$ ratios.\\
In Fig.6, for energy and Fig.7 for angular distributions, 
we show the resulting effects in the unpolarized $W^-$ ratios,
with $W^-_T$ and pure $W^-_L$ (up), or with $W^-_T$ and $G^-$ (down) 
as suggested by the CSMG equivalence.\\
The comparison of the middle and down figures shows how the CSMG hypothesis 
can be tested from its specific behaviours, with larger energy decreases
than in the CSM violating cases.\\

\underline{Effects of pure b compositeness}\\

The same illustrations are made in Fig.8,9,10 with the effects of 
$b_{L,R}$ compositeness. As expected only the effects of $b_{L}$ compositeness
are significant and essentially no effect of an effective bottom mass
can be observed.\\

\underline{Effects of both t and b compositeness}\\

Finally in Fig.11,12,13 we show the resulting modifications appearing when both
$t_{L,R}$ and $b_{L,R}$ compositeness are introduced.\\
The comparison with the two above cases (pure $t$ and pure $b$)  
shows different behaviours. Globally the ratios are weaker than the ones
due to pure $t$ or pur $b$ compositeness because of a better factorization
of the form factor effects preserving the SM combinations.\\

\section{Summary}

We have computed the cross sections of the $gb\to \gamma b, Hb, Zb, W^-t$
processes with point-like couplings of $\gamma , H, Z, W^-$
to top or bottom quarks modified by the introduction of specific form factors 
suggested by $t_L$, $t_R$, $b_L$, $b_R$ compositeness.\\ 
We also looked at the possible
effects of s-dependent effective top or bottom masses $m_f(s)$.\\
We treated separately the transverse and the longitudinal gauge boson
production.
In the longitudinal case we have compared the crude results due to the 
introduction of form factors in the gauge boson couplings to those 
suggested by the CSMG assumption 
which assumes an effective equivalence with 
the Goldstone bosons $G^{0,-}$ amplitudes including now form factors in the 
Goldstone couplings.\\
We have given illustrations for the ratios of modified cross sections 
over standard ones. Specific modifications of the energy and angular dependences
of these ratios are produced depending on the location of the form 
factors.
So interesting informations about compositeness and the CSM concept
should be obtained from the measurements of these ratios.\\
They should confirm
the corresponding results that would be obtained from other proceeses 
involving Higgs boson, top and bottom quarks in $e^+e-$, in photon-photon
and in hadronic collisions,\cite{CSMrev,CSMbot}.\\
The observability of such processes can be for example found in
\cite{Moortgat,Denterria,Craig} for $e^+e^-$, \cite{Contino,Richard} for proton-proton
and  \cite{gammagamma} for photon-photon.\\

\newpage

\begin{figure}[p]
\[
\epsfig{file=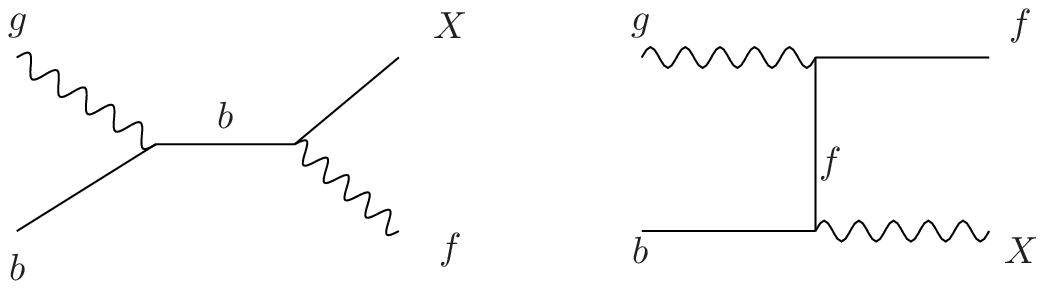, height=4.cm}
\]\\
\vspace{-1cm}
\caption[1]  {Born diagrams for $gb\to Xf$; $Xf\equiv Hb, ~Zb, ~W^-t$.}

\end{figure}

\begin{figure}[p]

\[
\epsfig{file=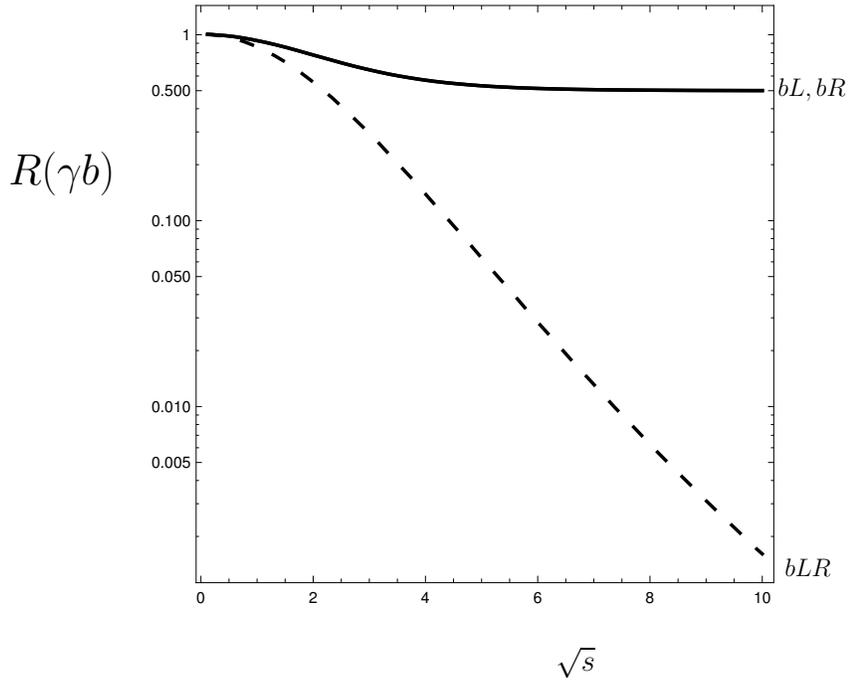, height=9.cm}
\]\\

\vspace{-1cm}
\caption[1]  {Ratios for  $gb\to \gamma b$ 
with $b_L$, $b_R$ compositeness or both.}

\end{figure}

\clearpage

\begin{figure}[p]

\[
\epsfig{file=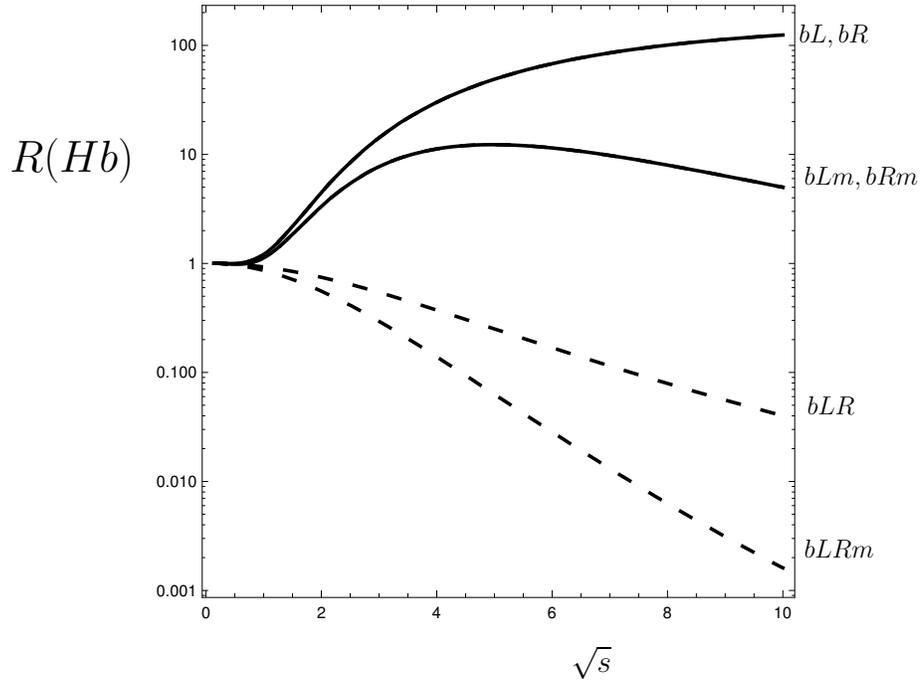, height=9.cm}
\]\\
\[
\epsfig{file=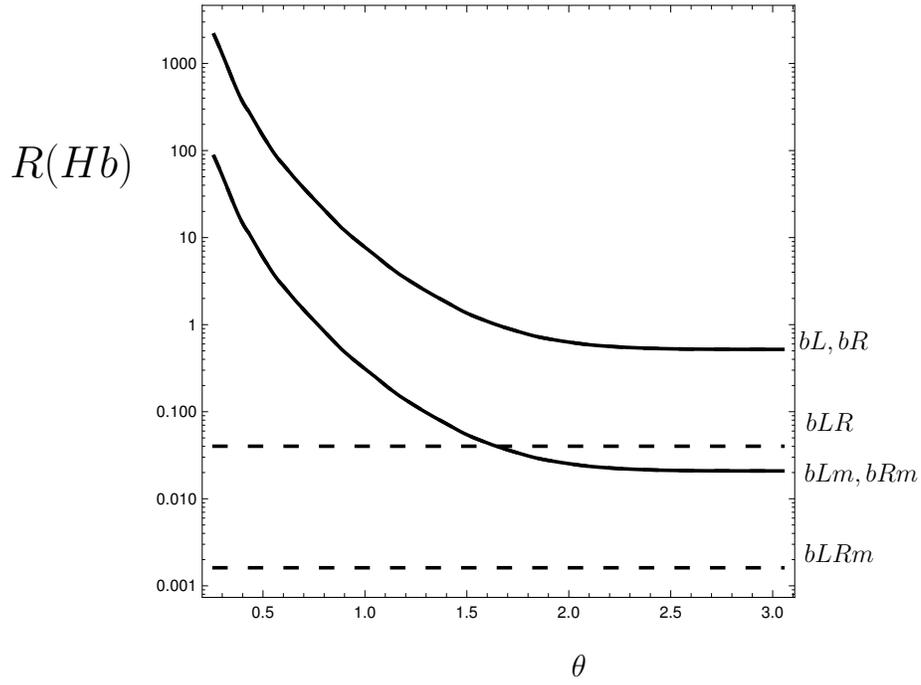, height=9.cm}
\]\\
\vspace{-1cm}
\caption[1]  {Ratios for  $gb\to H b$, energy dependence (up), 
angular dependence at 10 TeV (down),
with $b_L$, $b_R$ compositeness or both.}

\end{figure}

\clearpage

\begin{figure}[p]
\[
\epsfig{file=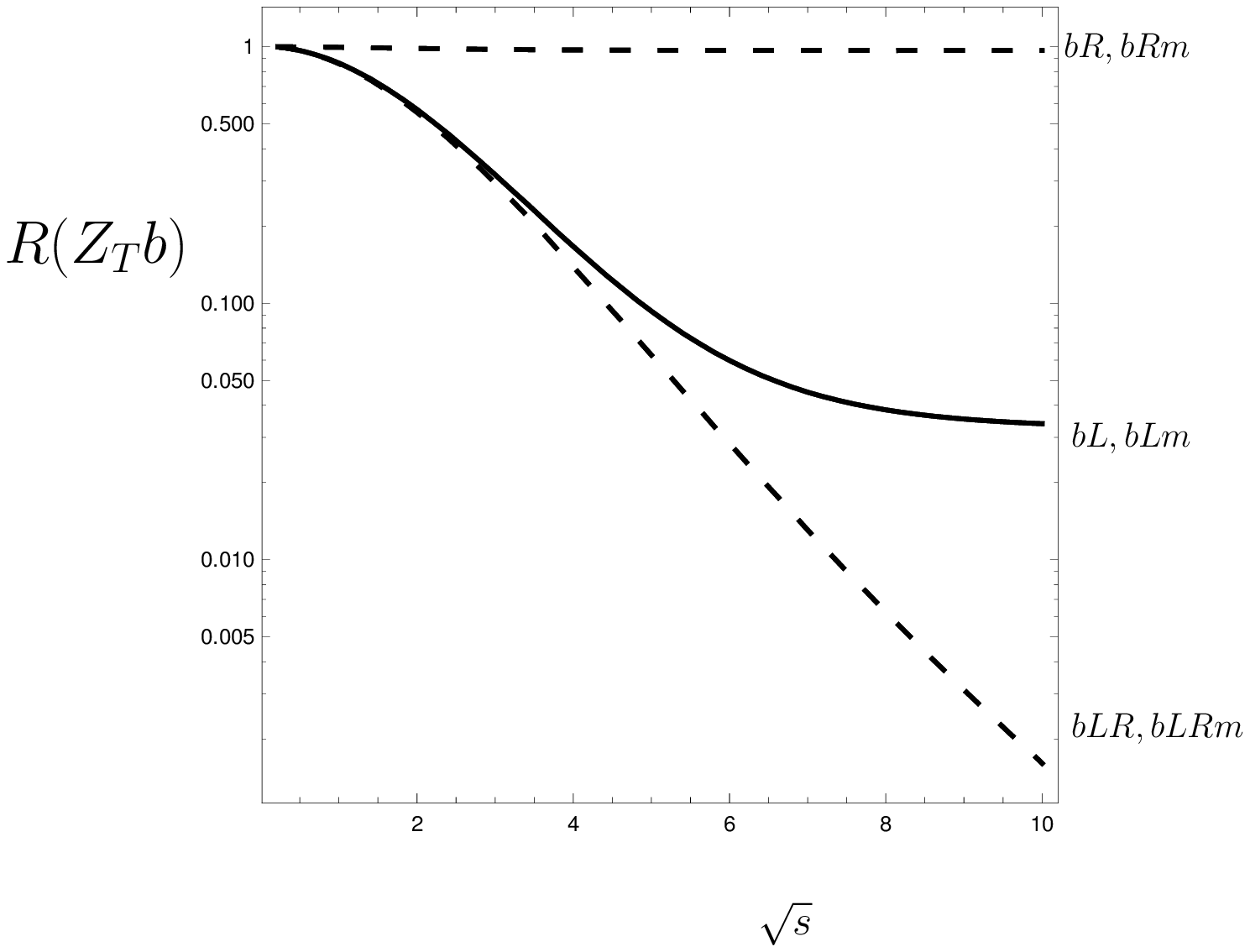, height=6.cm}
\]\\
\[
\epsfig{file=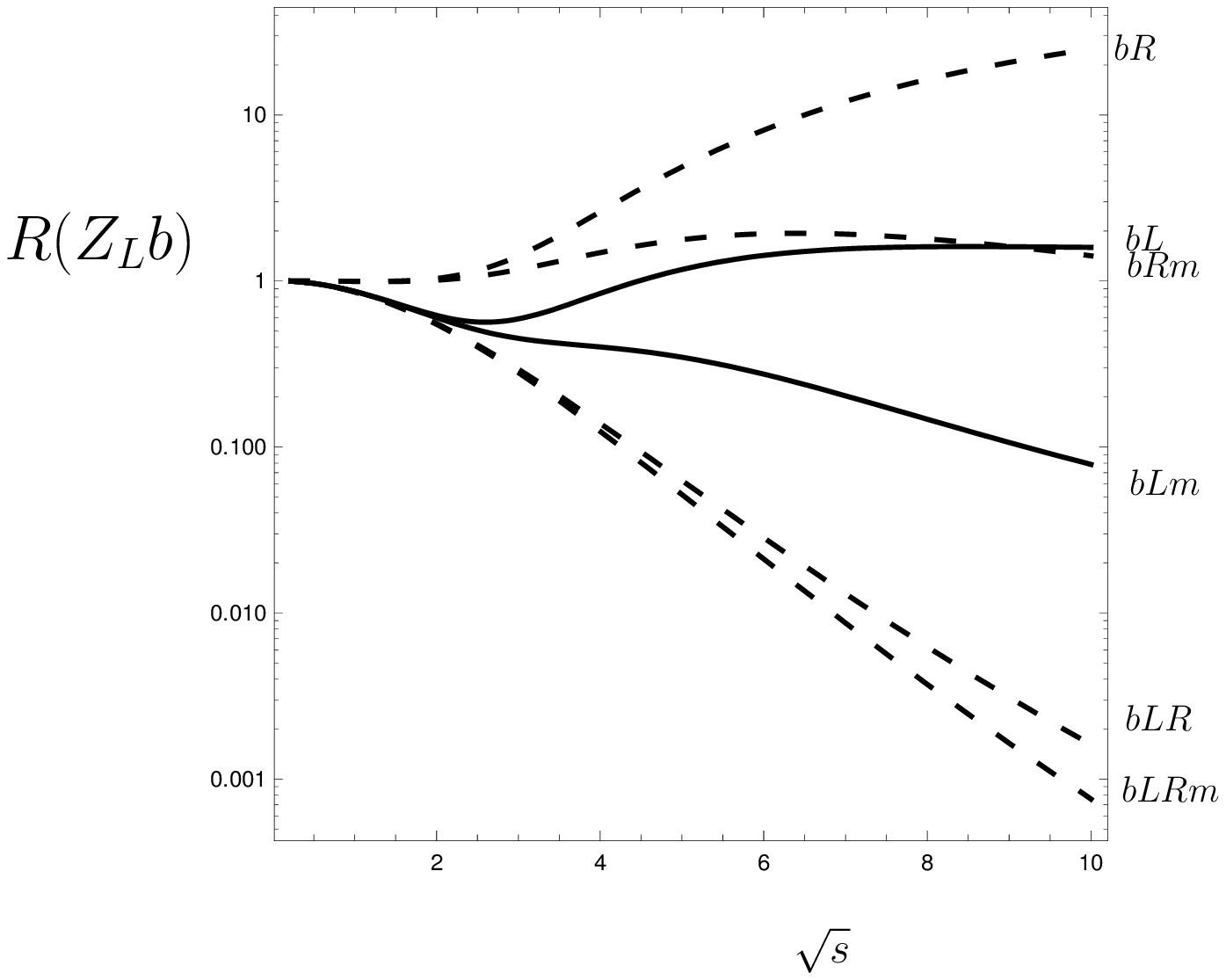, height=6.cm}
\]\\
\[
\epsfig{file=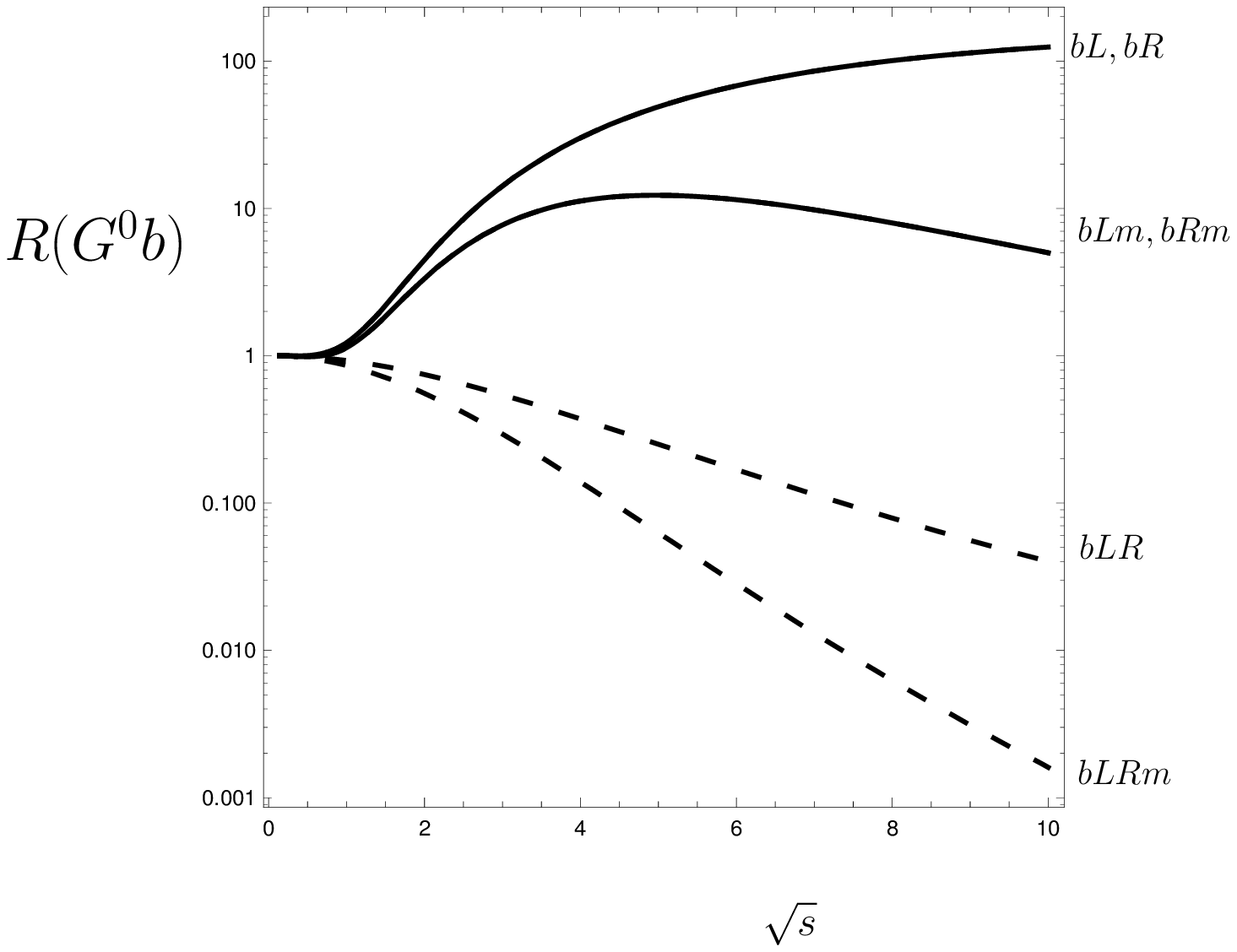, height=6.cm}
\]\\
\vspace{-1cm}
\caption[1]  {Ratios for  $gb\to Zb$, with $b_L$, $b_R$ compositeness or both.}

\end{figure}

\clearpage

\begin{figure}[p]
\[
\epsfig{file=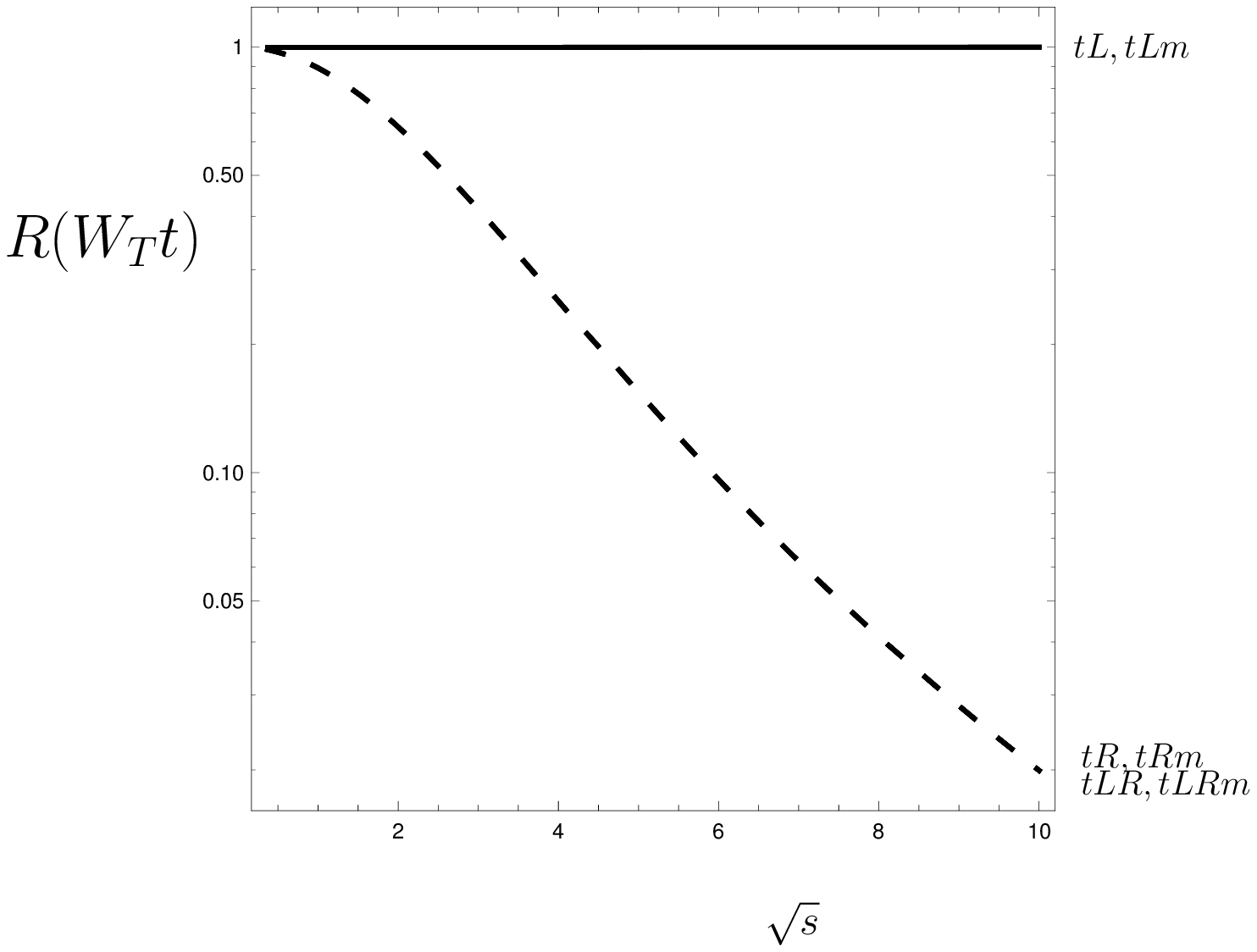, height=6.cm}
\]\\
\[
\epsfig{file=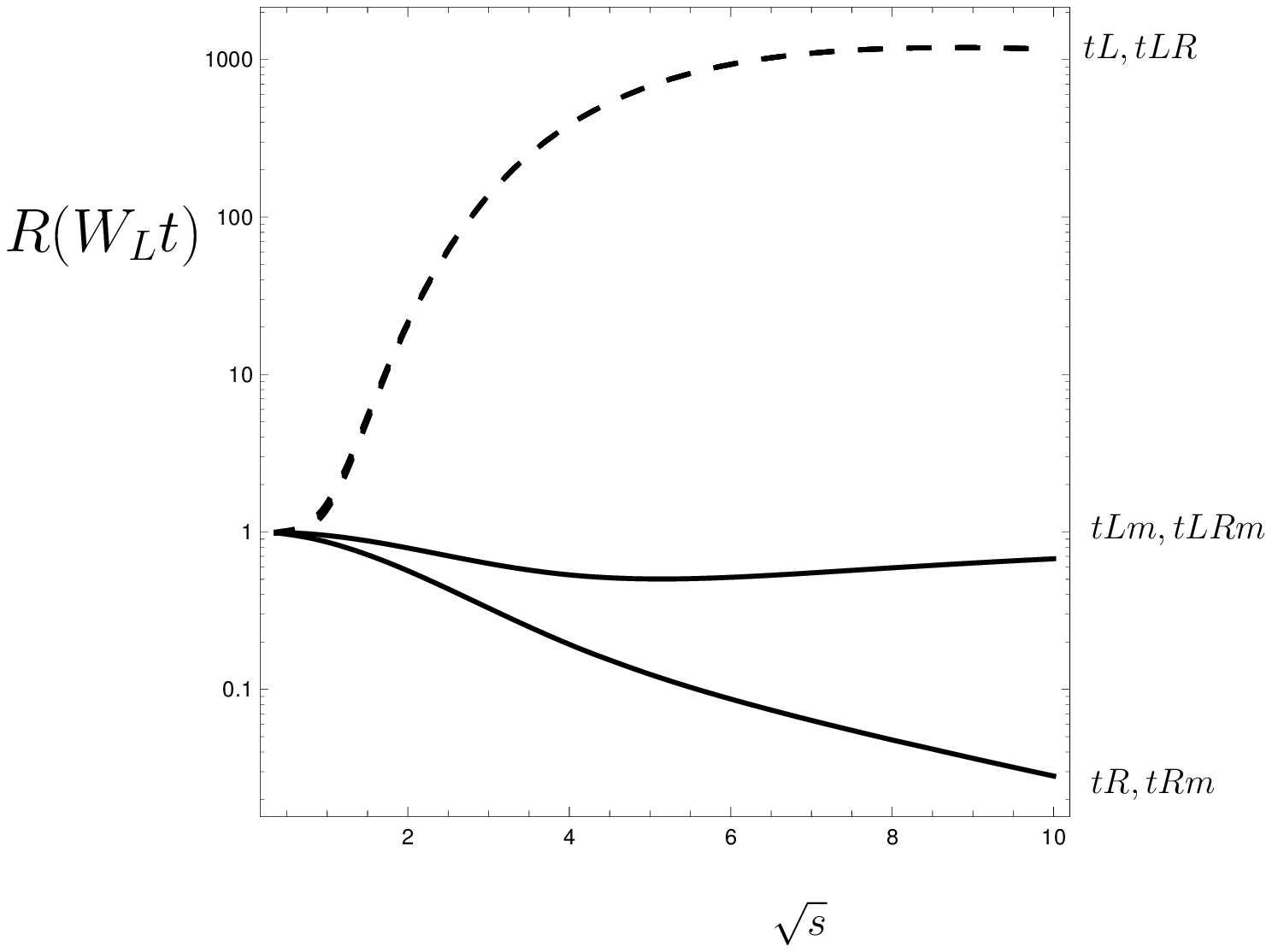, height=6.cm}
\]\\
\[
\epsfig{file=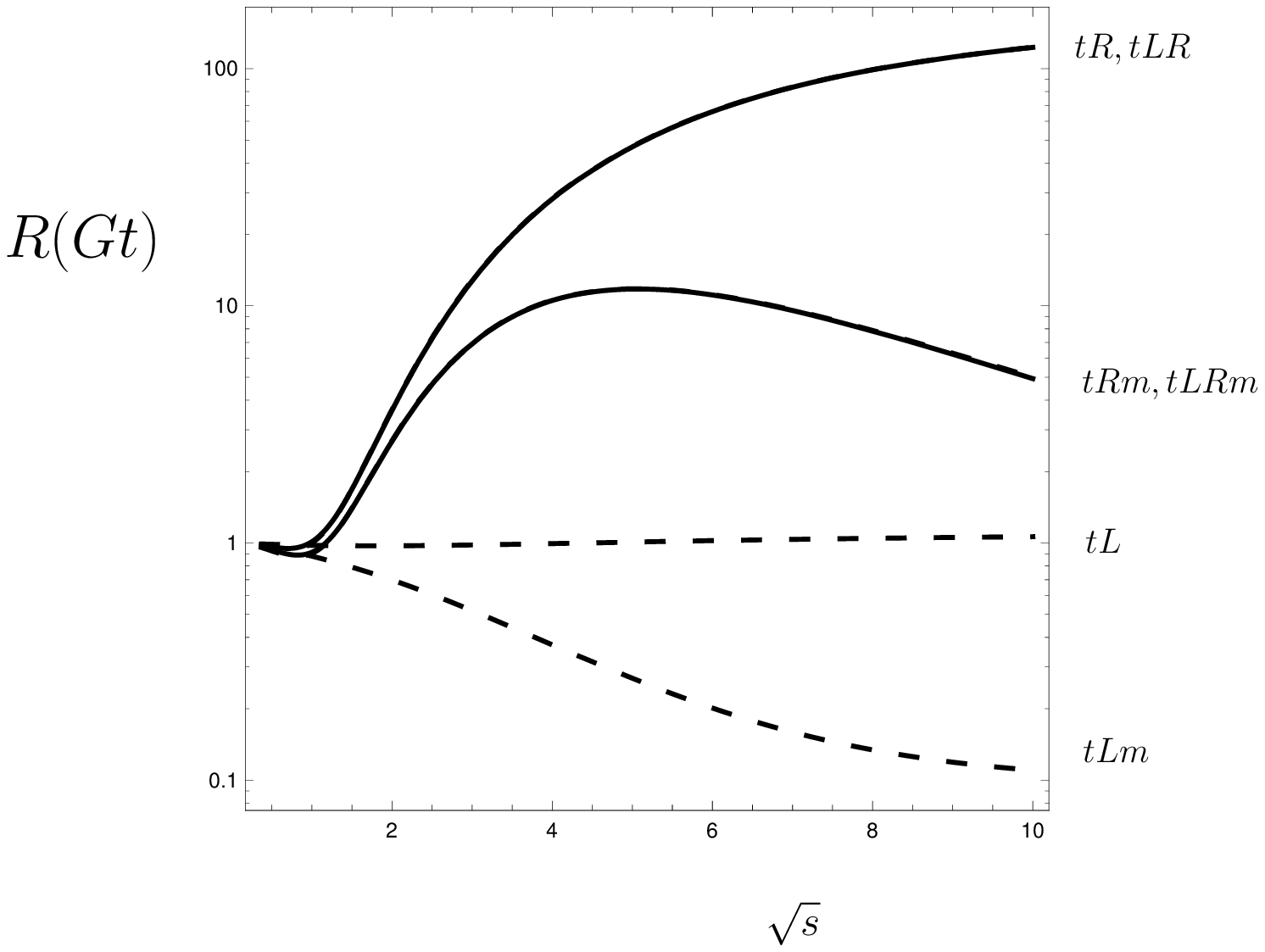, height=6.cm}
\]\\
\vspace{-1cm}
\caption[1] {Ratios for  $gb\to W^- t$, with $t_L$, $t_R$ compositeness or both.}

\end{figure}

\clearpage

\begin{figure}[p]
\[
\epsfig{file=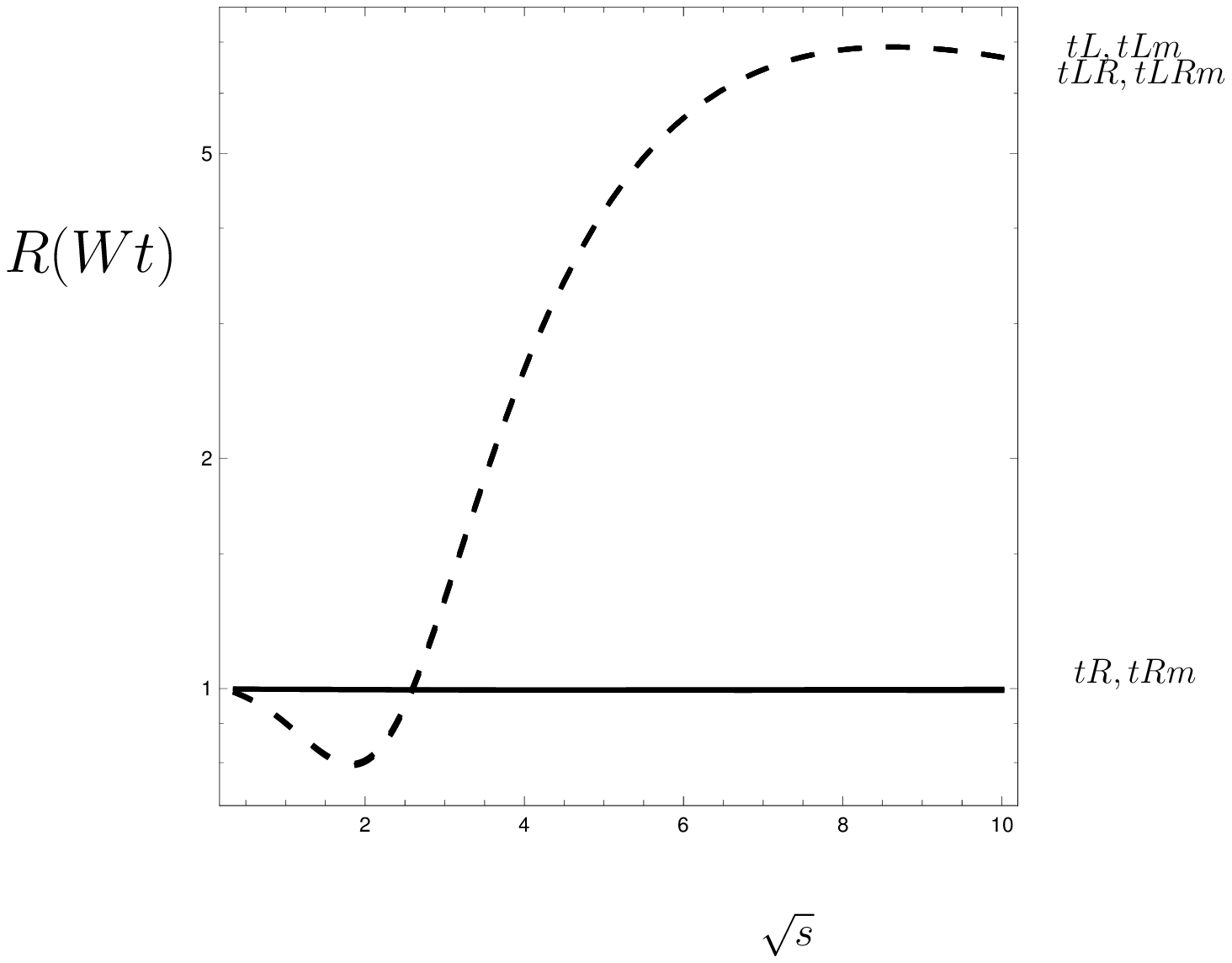, height=9.cm}
\]\\
\[
\epsfig{file=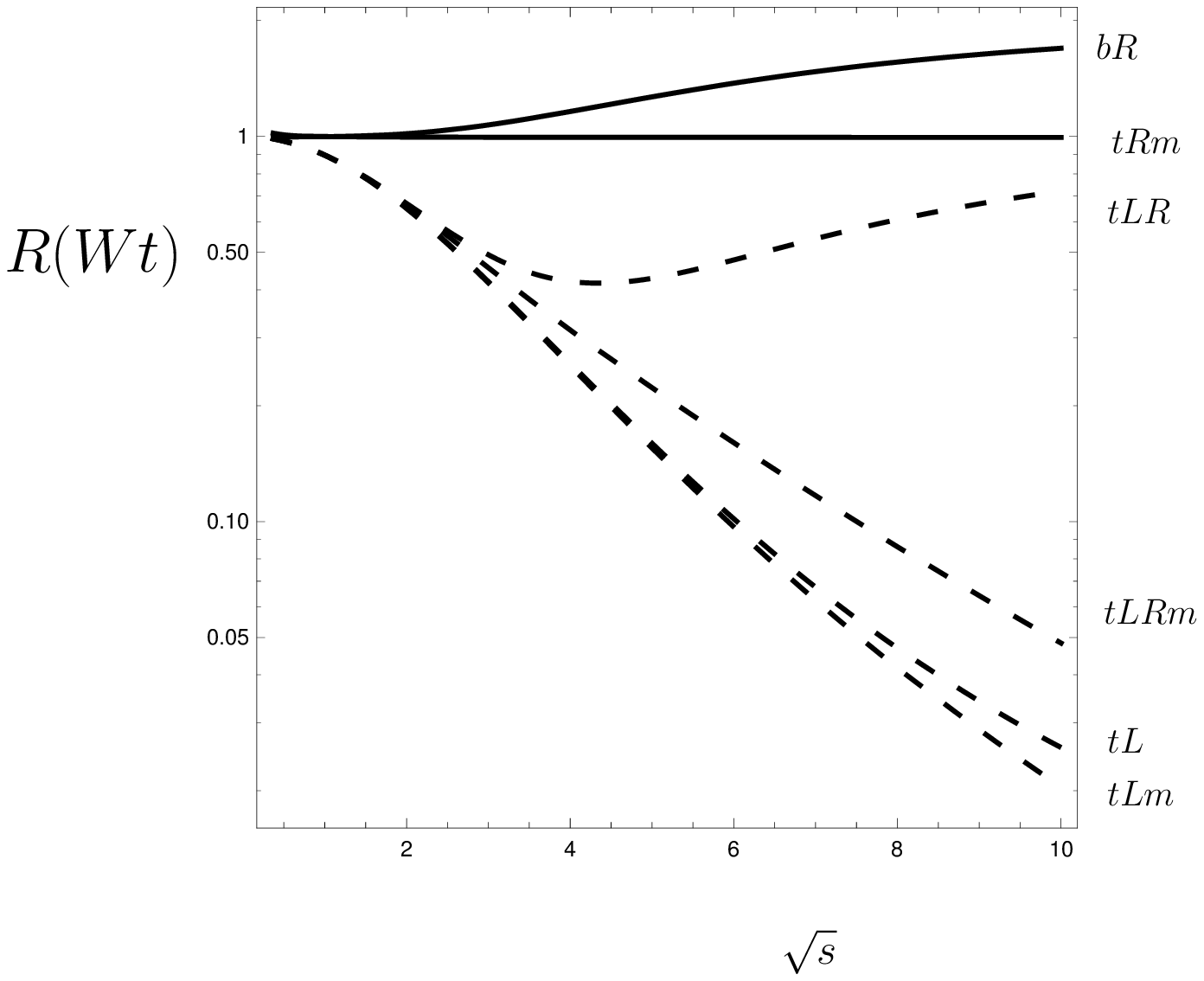, height=9.cm}
\]\\

\vspace{-1cm}
\caption[1]  {Ratios for  $gb\to W^- t$, from $W_T+W_L$(up),
from $W_T+G$(down), with $t_L$, $t_R$ compositeness or both.}

\end{figure}

\clearpage

\begin{figure}[p]
\[
\epsfig{file=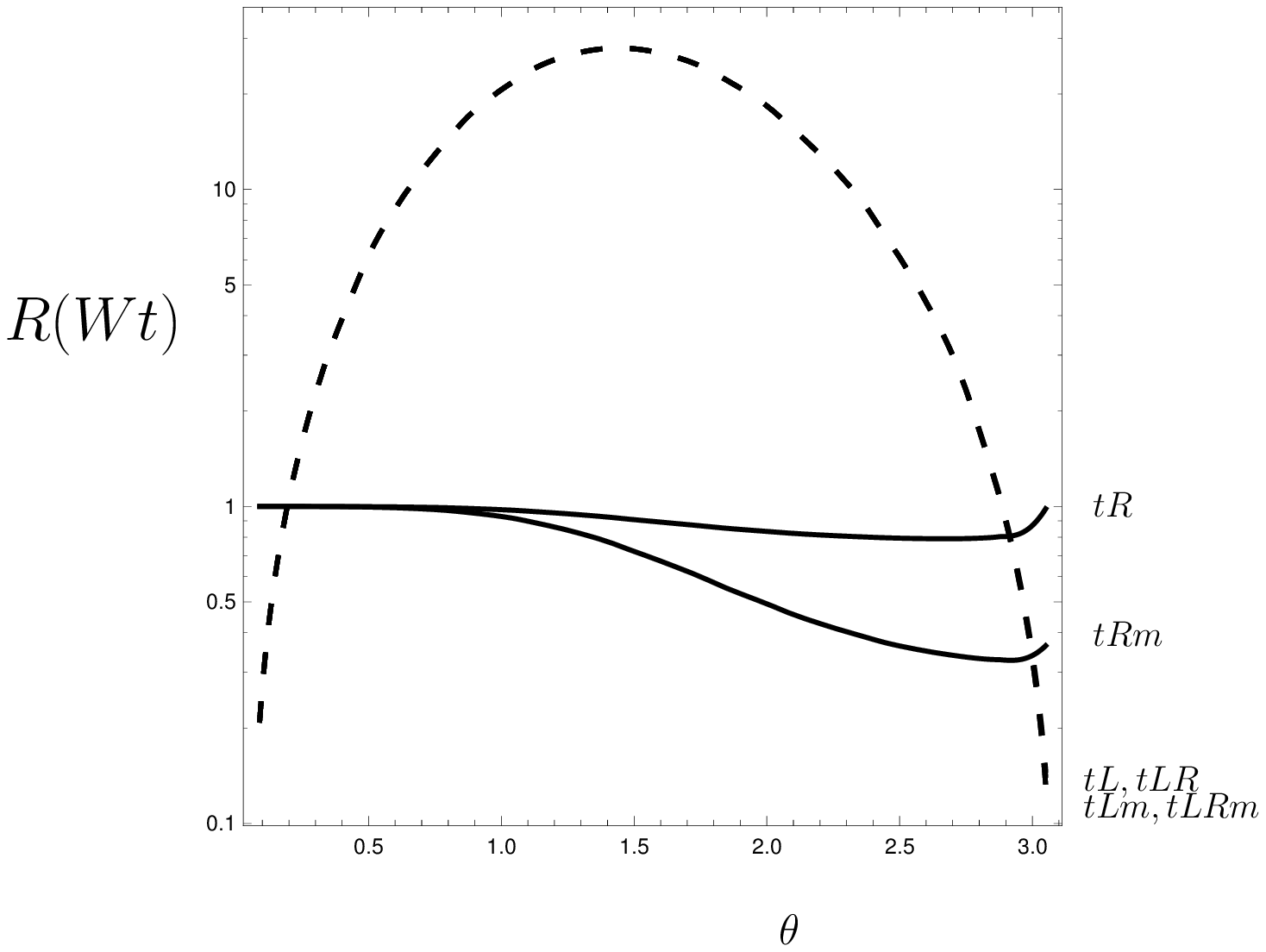, height=9.cm}
\]\\
\[
\epsfig{file=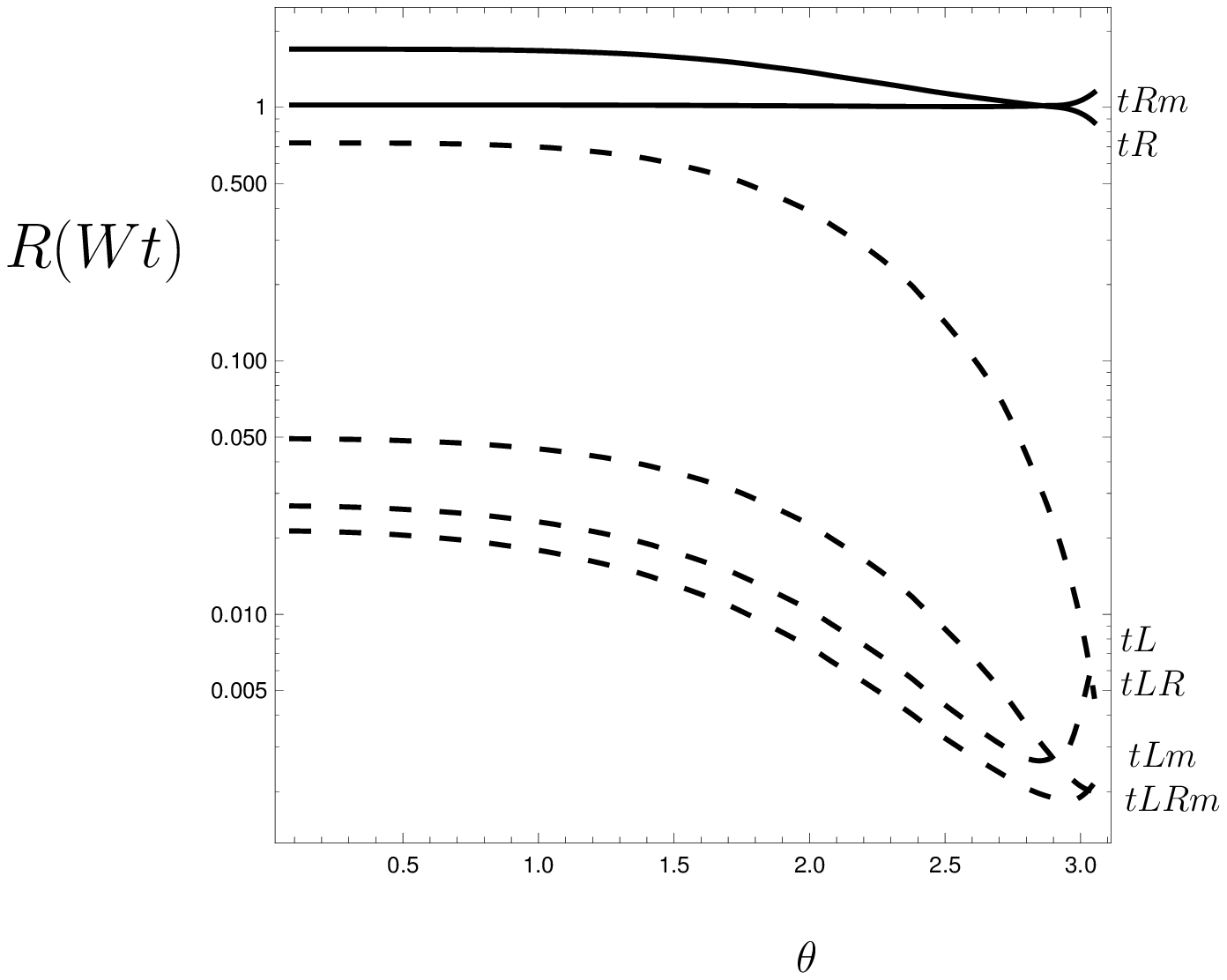, height=9.cm}
\]\\

\vspace{-1cm}
\caption[1]  {Angular distributions at 10 TeV of ratios for  $gb\to W^- t$, from $W_T+W_L$(up),
from $W_T+G$(down), with $t_L$, $t_R$ compositeness or both.}

\end{figure}

\clearpage

\begin{figure}[p]
\[
\epsfig{file=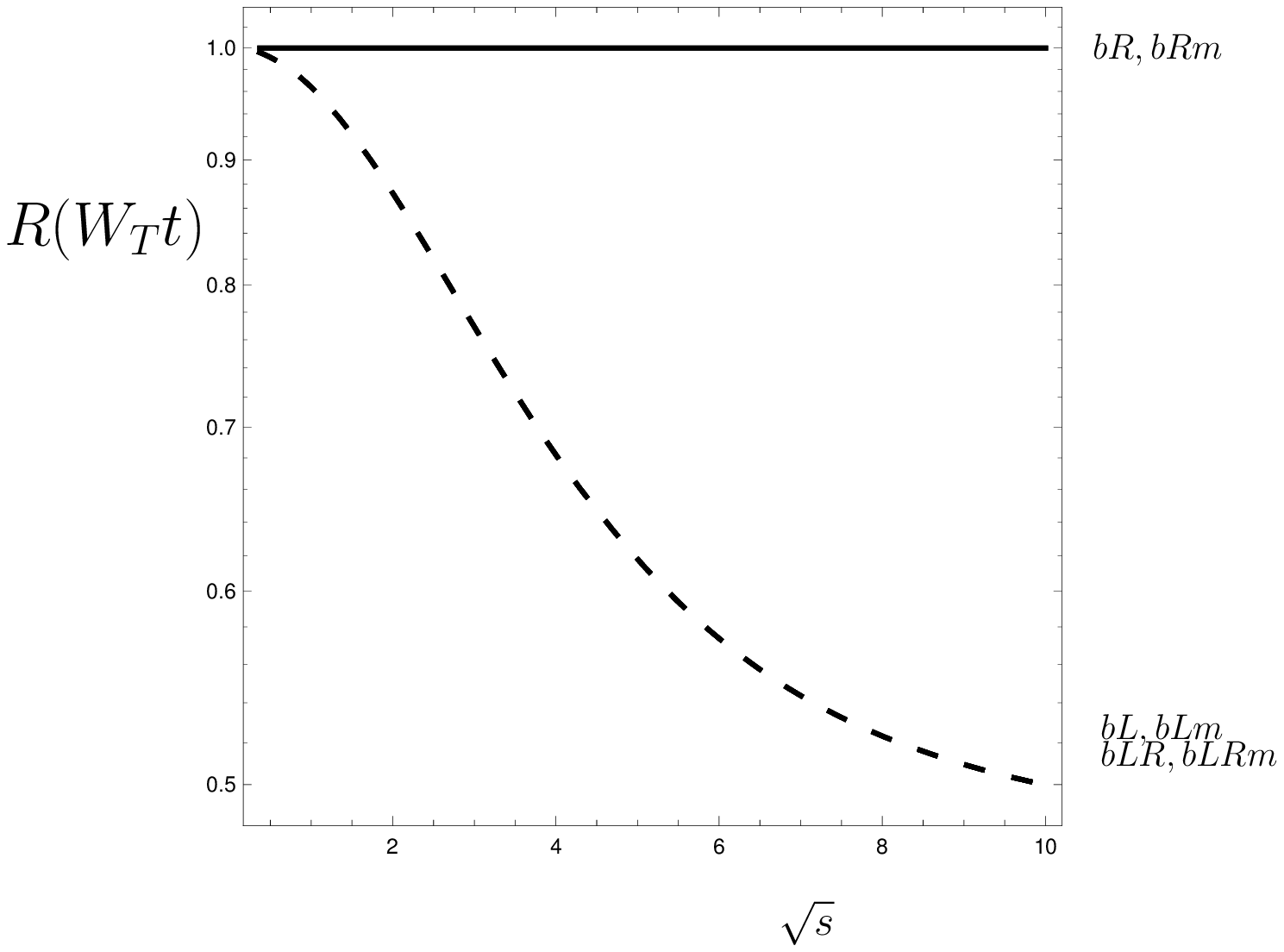, height=6.cm}
\]\\
\[
\epsfig{file=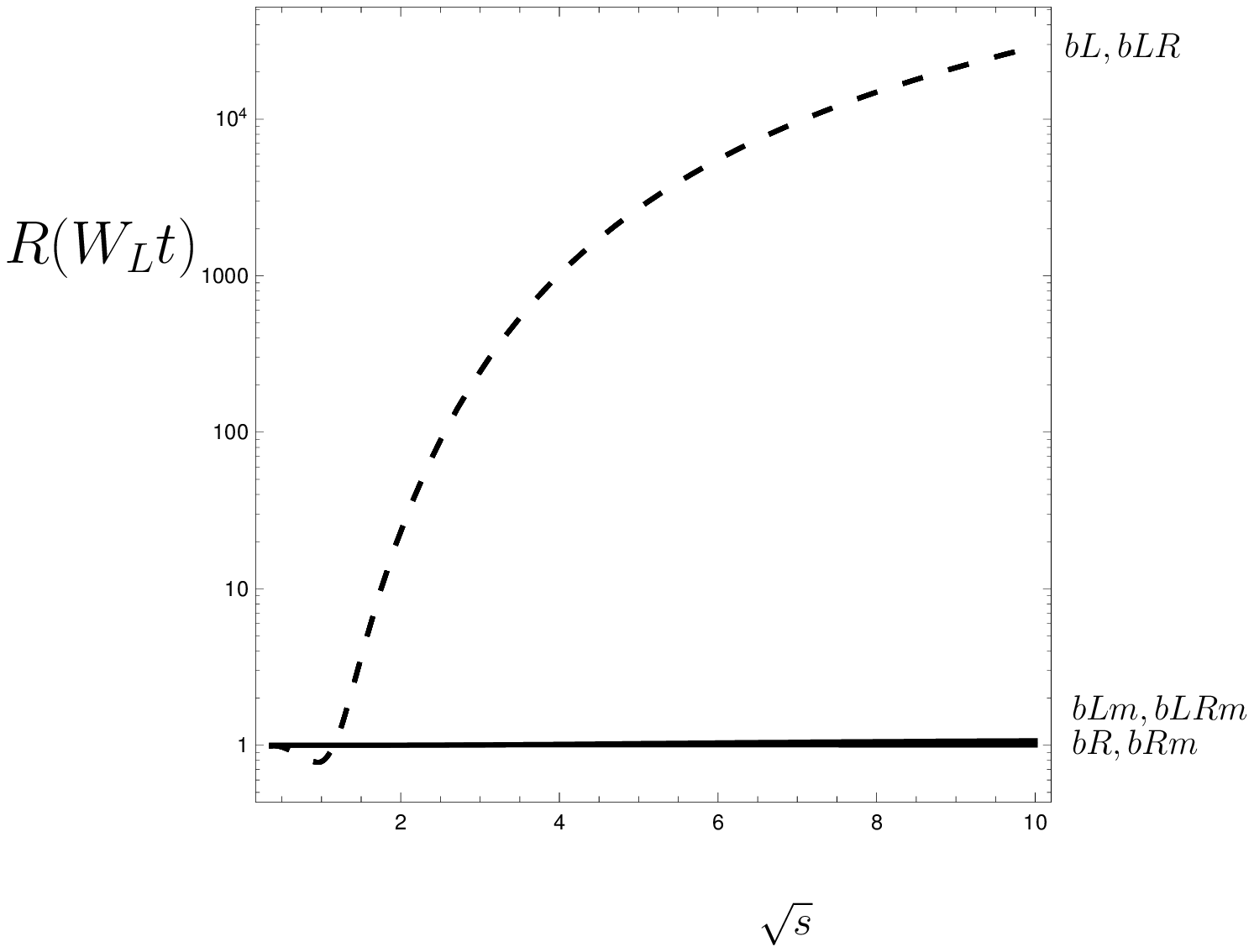, height=6.cm}
\]\\
\[
\epsfig{file=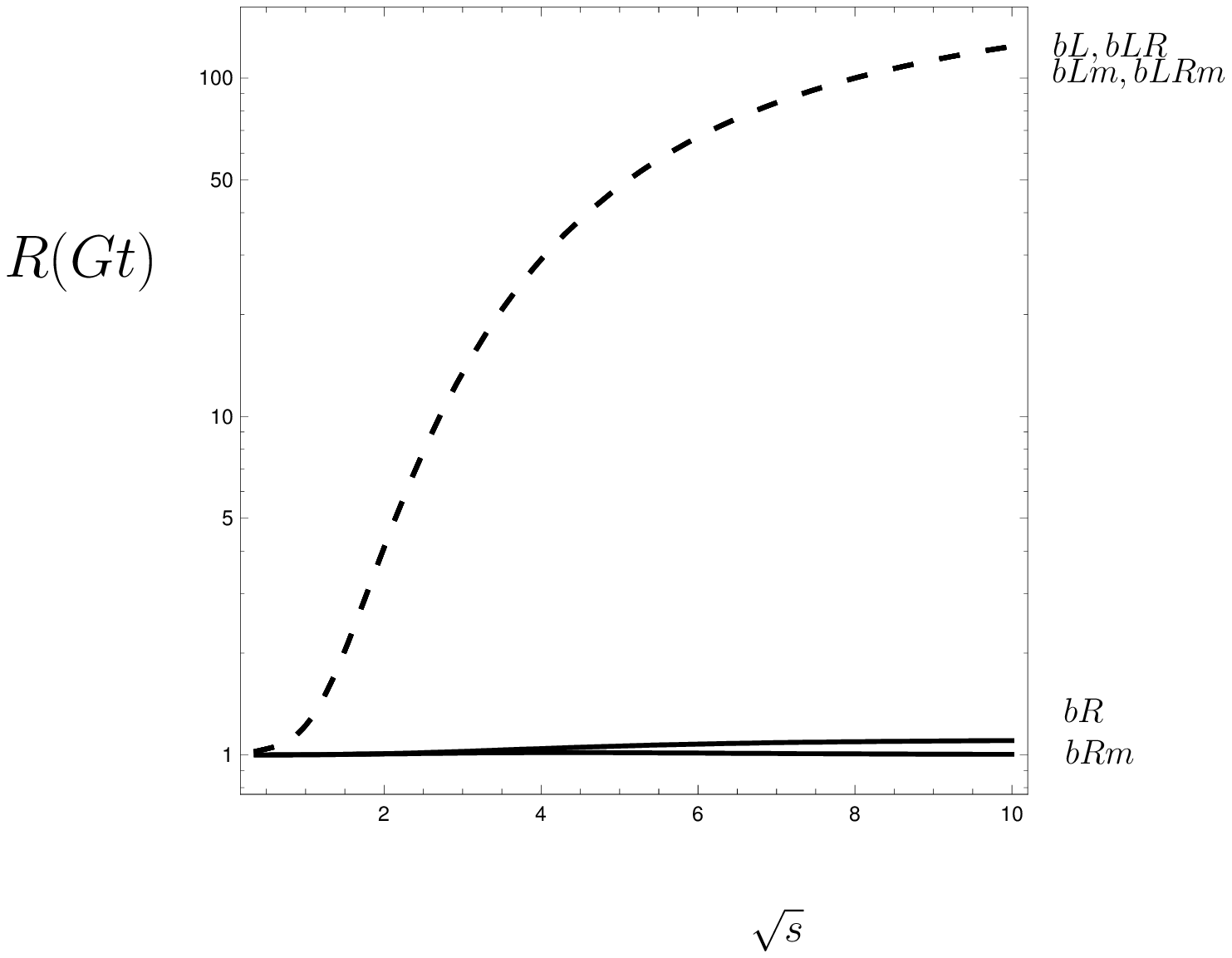, height=6.cm}
\]\\
\vspace{-1cm}
\caption[1]  {Ratios for  $gb\to W^- t$, with $b_L$, $b_R$ compositeness or both.}

\end{figure}

\clearpage

\begin{figure}[p]
\[
\epsfig{file=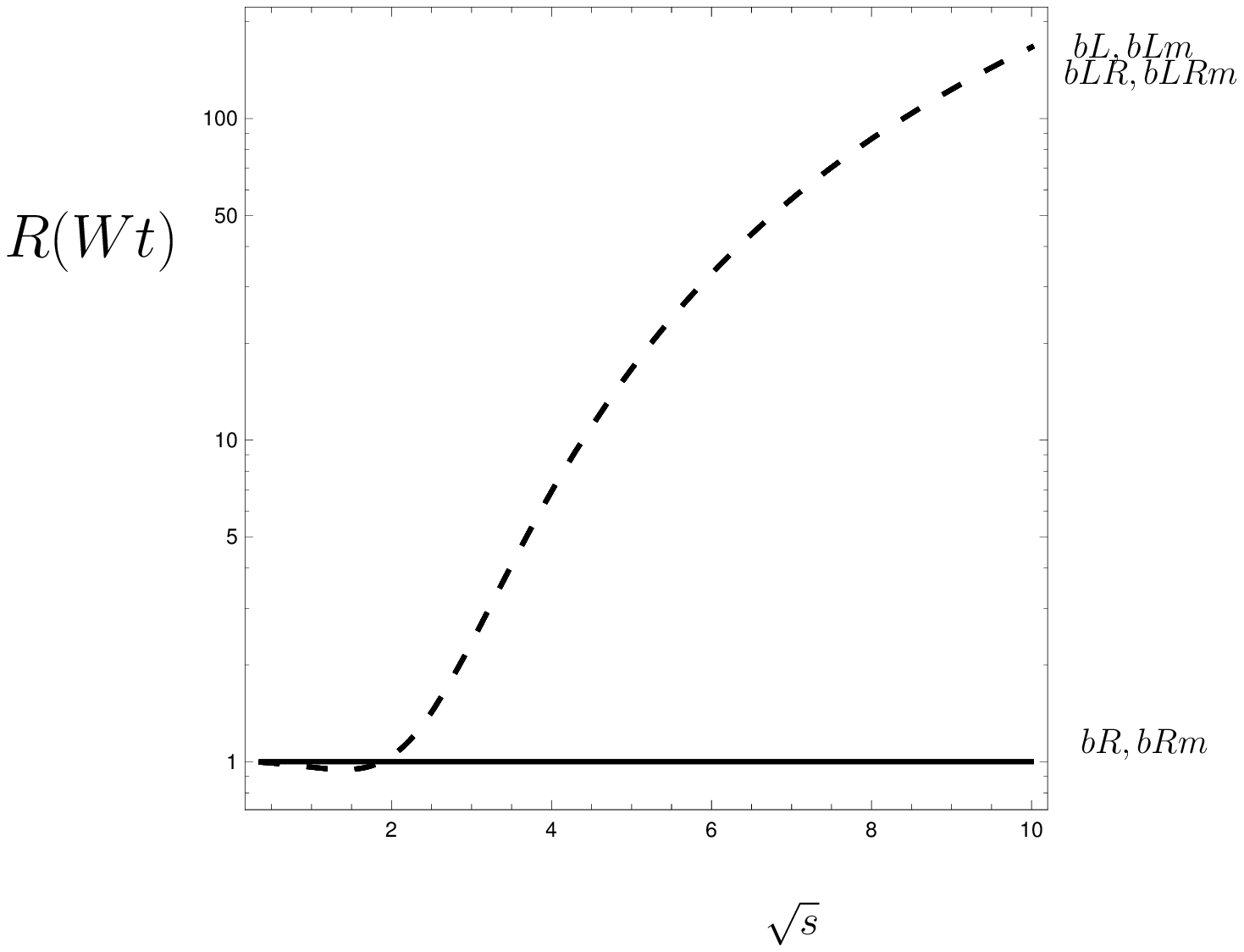, height=9.cm}
\]\\
\[
\epsfig{file=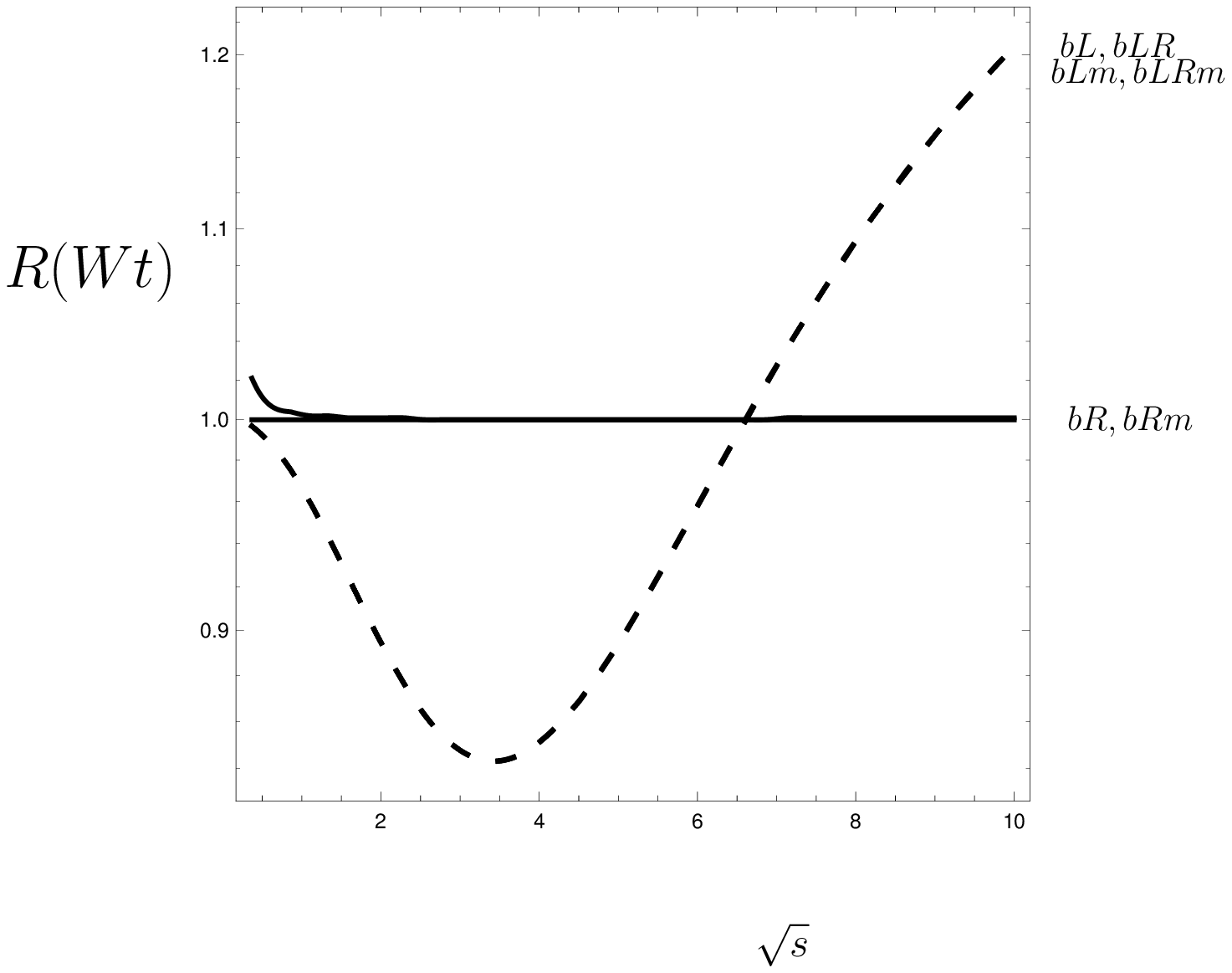, height=9.cm}
\]\\

\vspace{-1cm}
\caption[1]  {Ratios for  $gb\to W^- t$, from $W_T+W_L$(up),
from $W_T+G$(down), with $b_L$, $b_R$ compositeness or both.}

\end{figure}

\clearpage

\begin{figure}[p]
\[
\epsfig{file=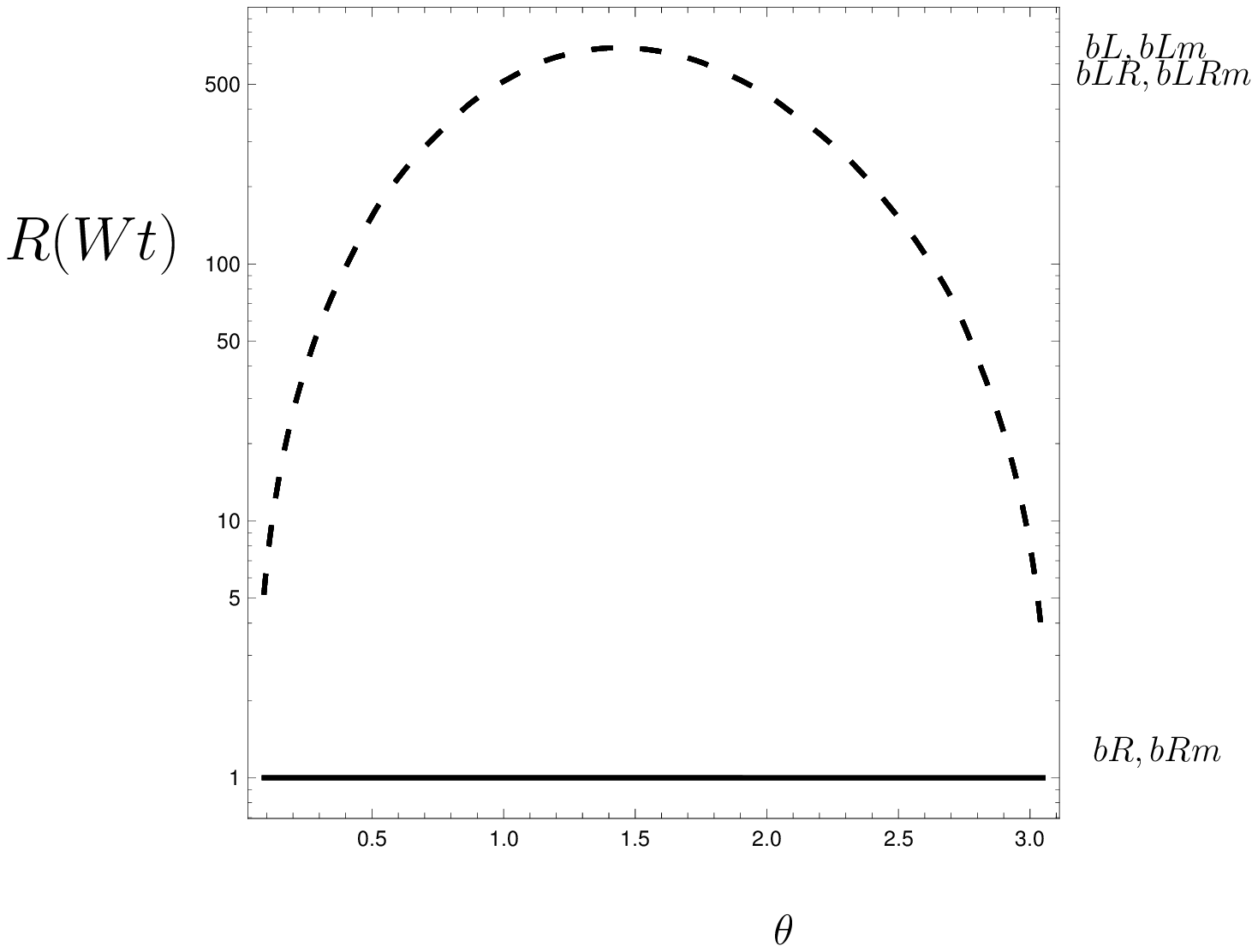, height=9.cm}
\]\\
\[
\epsfig{file=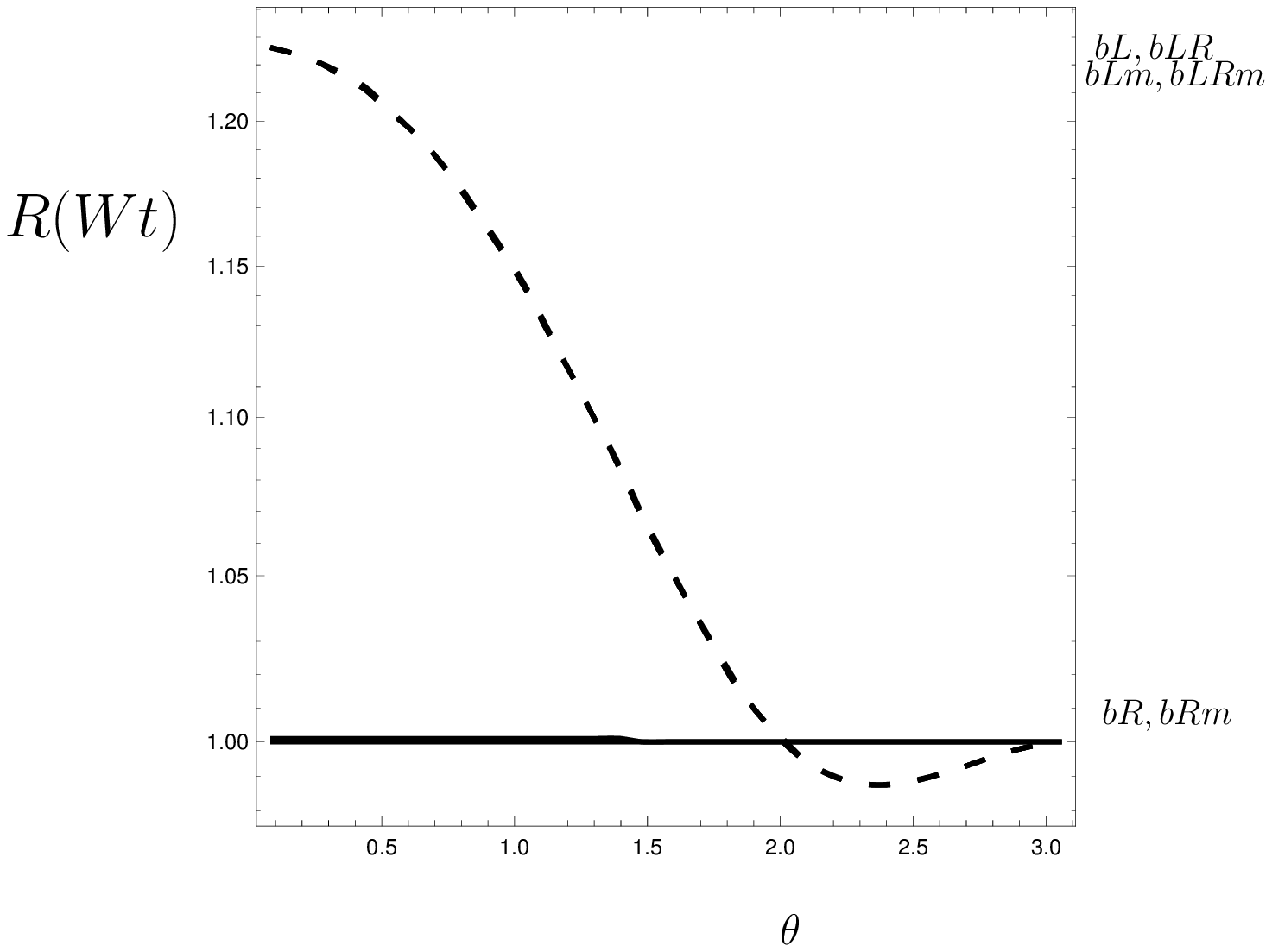, height=9.cm}
\]\\

\vspace{-1cm}
\caption[1]  {Angular distributions at 10 TeV of ratios for  $gb\to W^- t$, from $W_T+W_L$(up),
from $W_T+G$(down), with $b_L$, $b_R$ compositeness or both.}

\end{figure}

\clearpage

\begin{figure}[p]
\[
\epsfig{file=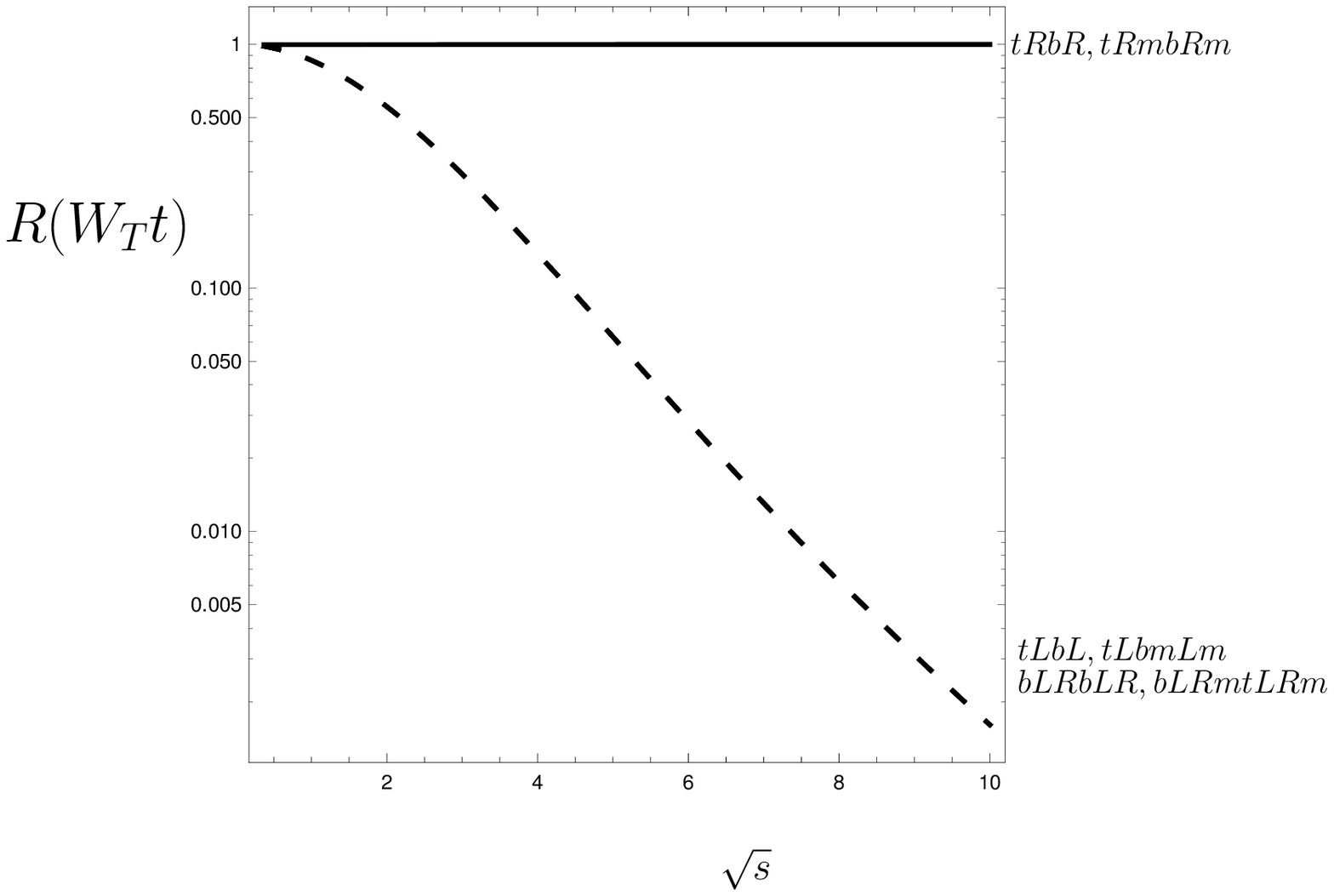, height=6.cm}
\]\\
\[
\epsfig{file=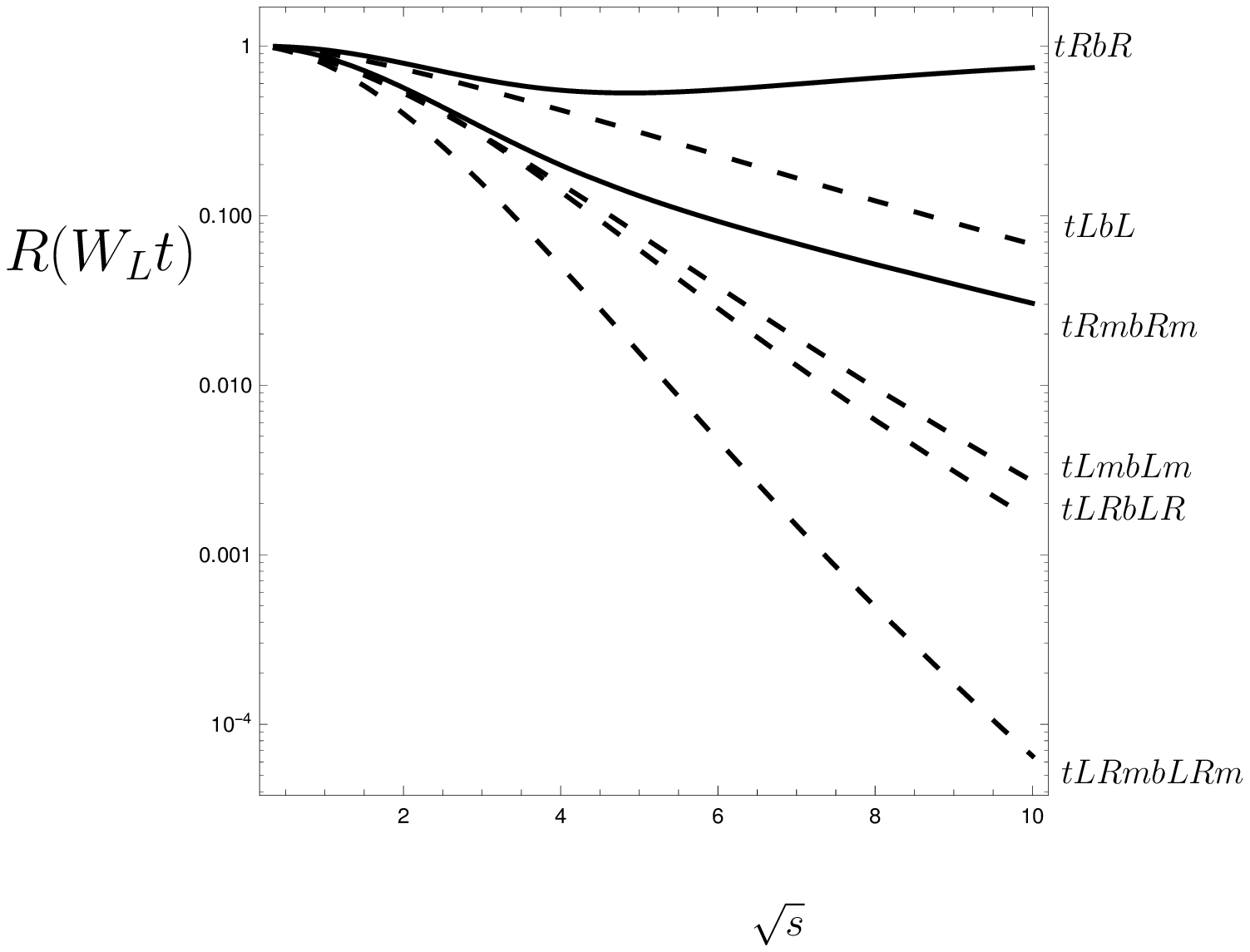, height=6.cm}
\]\\
\[
\epsfig{file=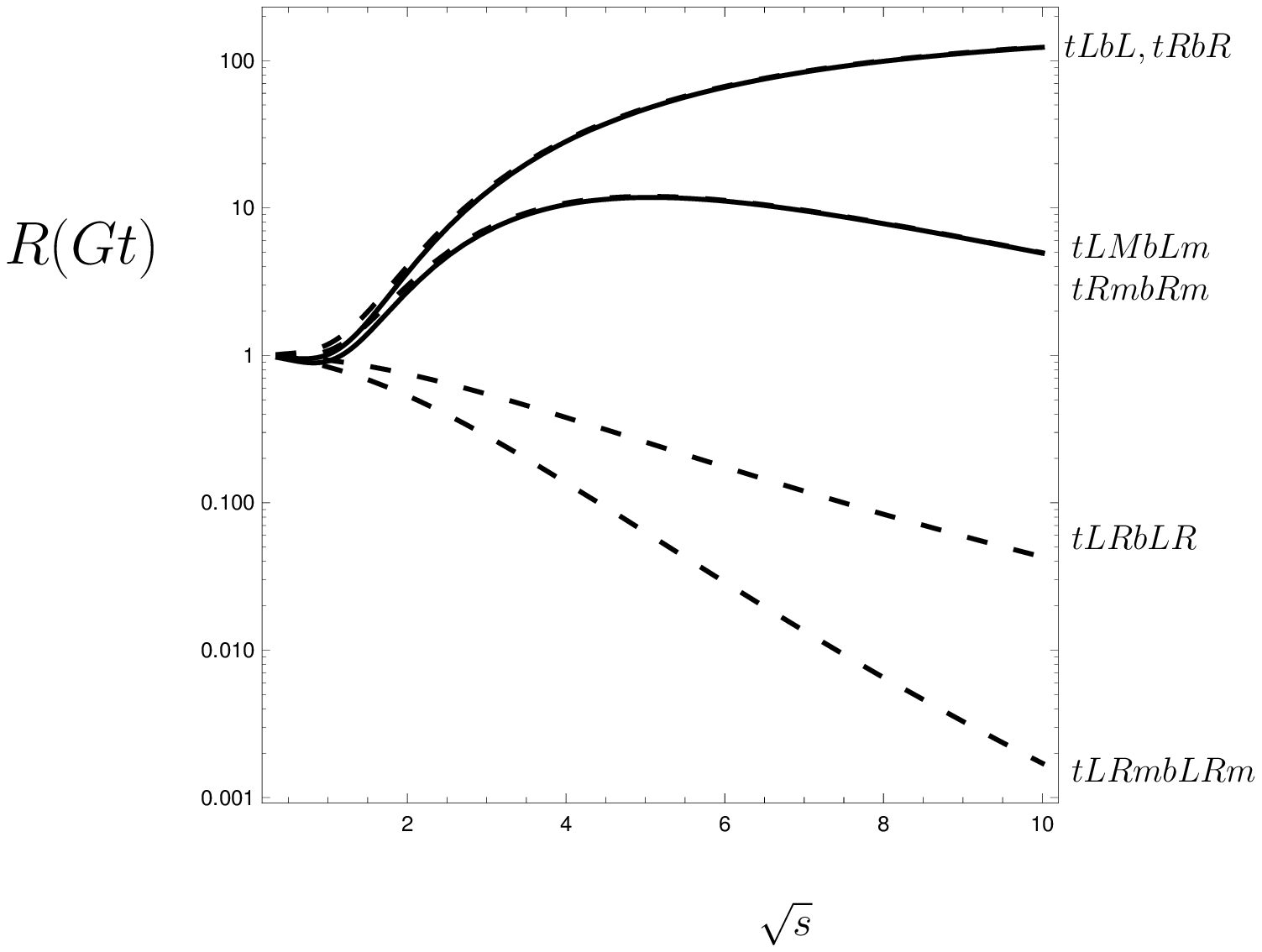, height=6.cm}
\]\\
\vspace{-1cm}
\caption[1]  {Ratios for  $gb\to W^- t$, with $t_L,b_L$ or $t_R,b_R$ compositeness or both.}

\end{figure}

\clearpage

\begin{figure}[p]
\[
\epsfig{file=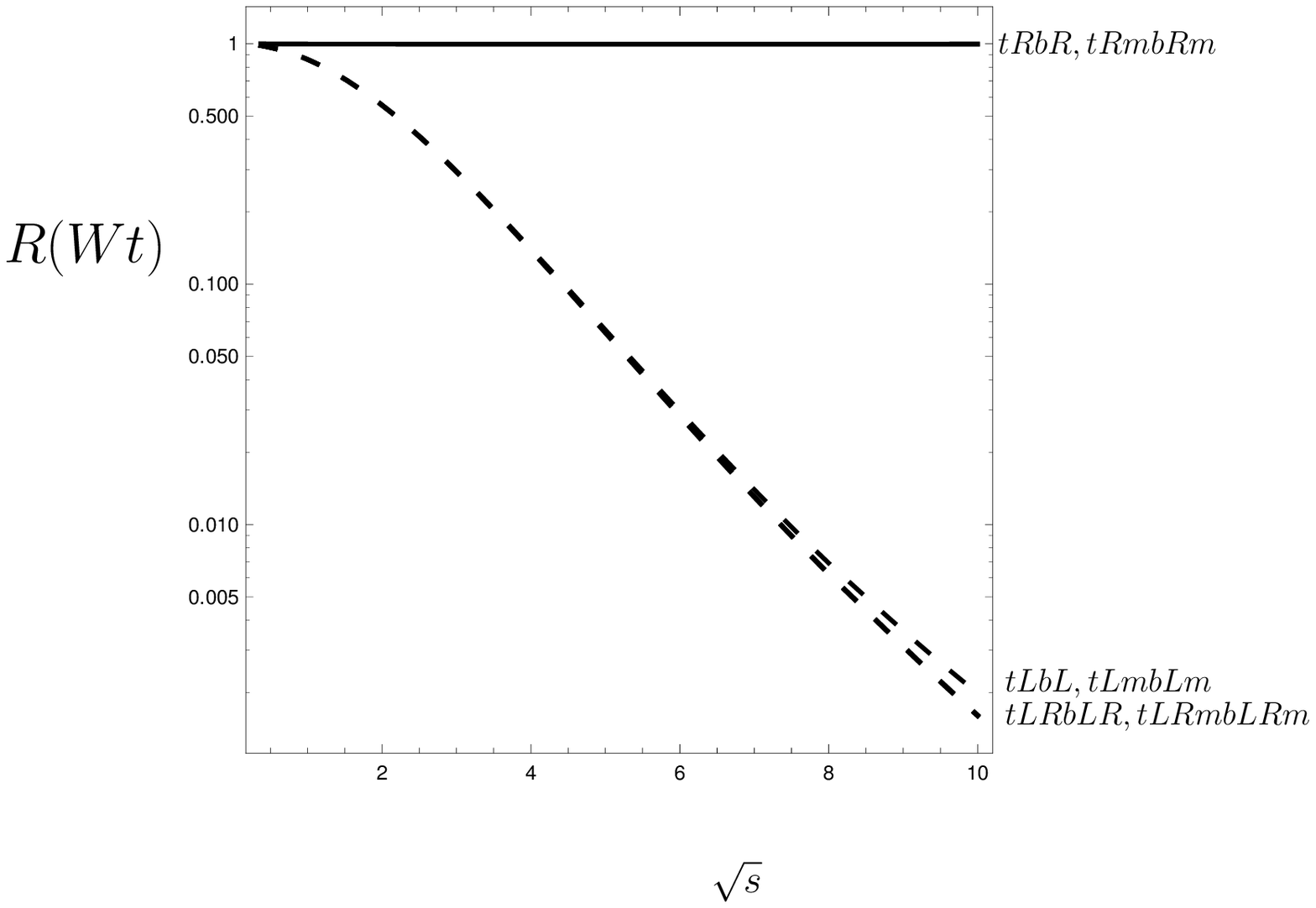, height=9.cm}
\]\\
\[
\epsfig{file=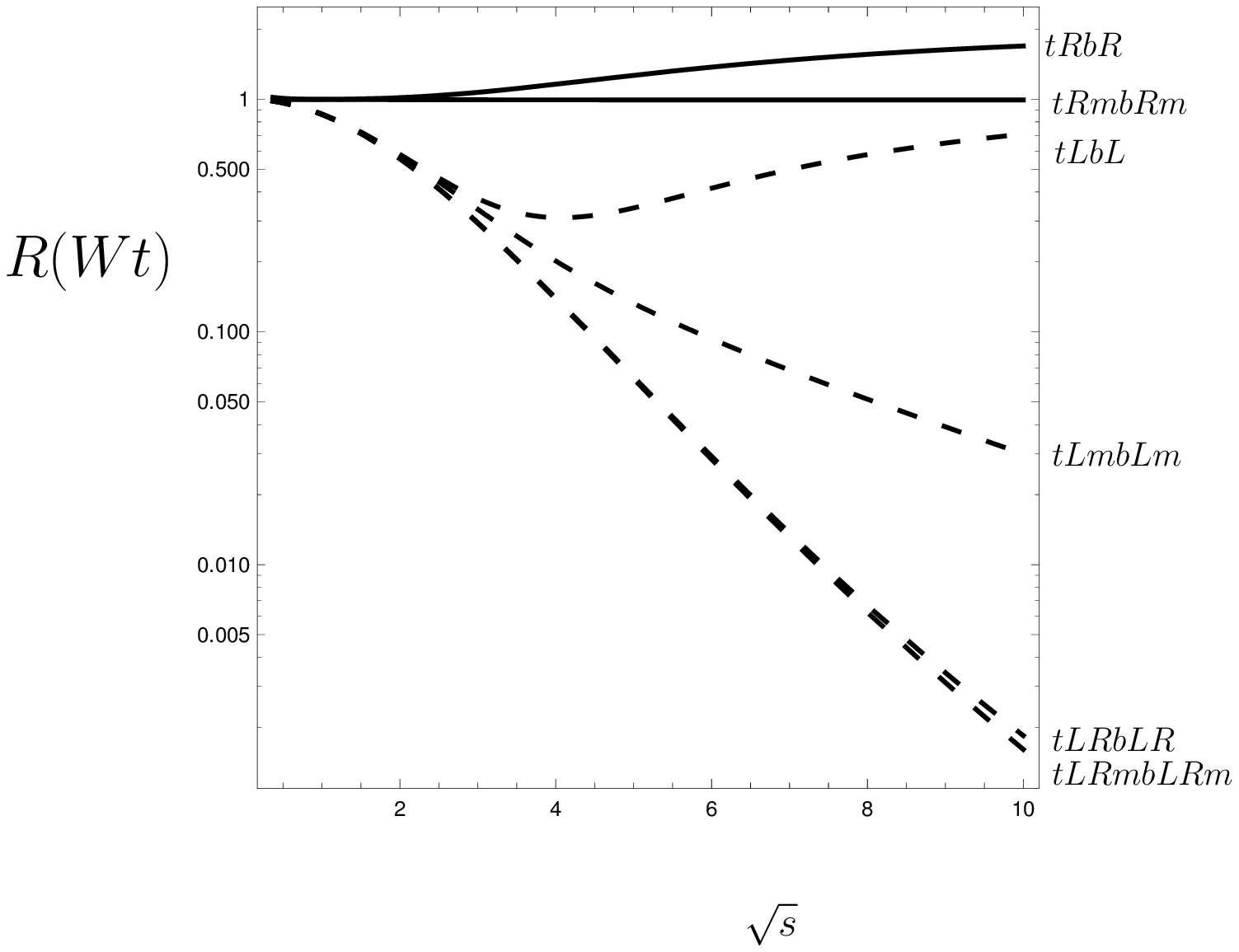, height=9.cm}
\]\\

\vspace{-1cm}
\caption[1]  {Ratios for  $gb\to W^- t$, from $W_T+W_L$(up),
from $W_T+G$(down), with $t_L,b_L$ or $t_R,b_R$ compositeness or both.}

\end{figure}

\clearpage

\begin{figure}[p]
\[
\epsfig{file=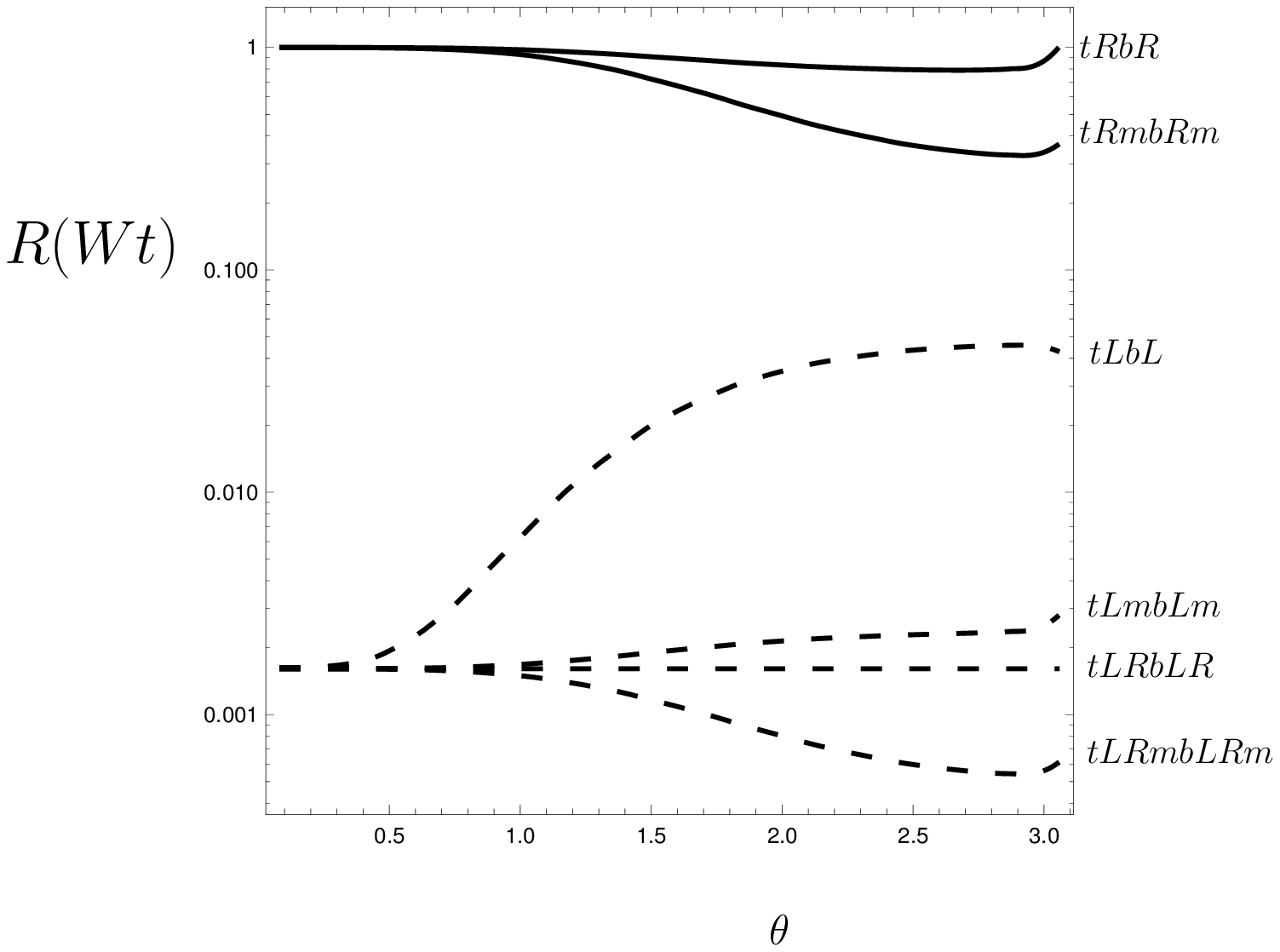, height=9.cm}
\]\\
\[
\epsfig{file=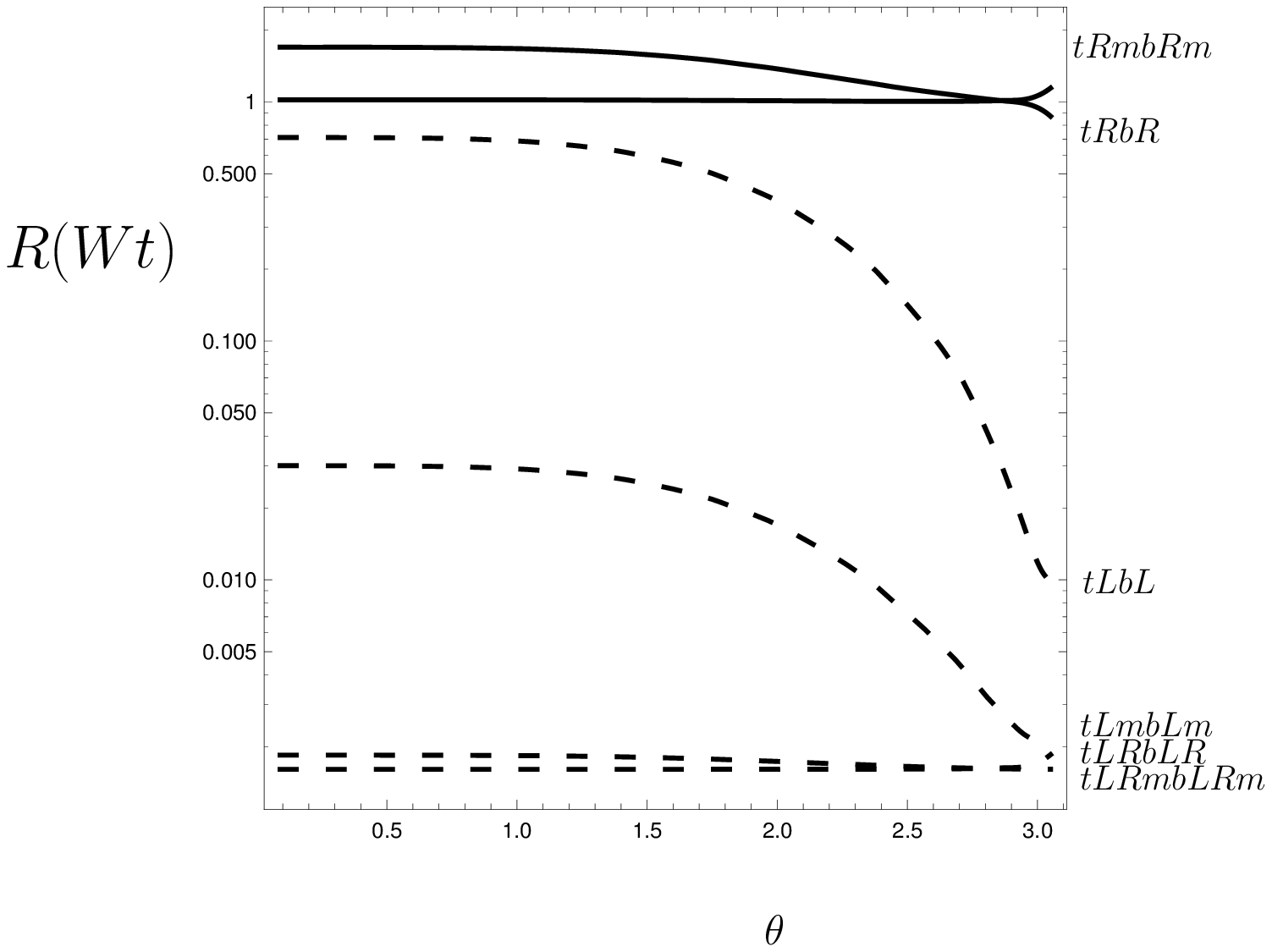, height=9.cm}
\]\\

\vspace{-1cm}
\caption[1]  {Angular distribution at 10 TeV of ratios for  $gb\to W^- t$, from $W_T+W_L$(up),
from $W_T+G$(down), with $t_L,b_L$ or $t_R,b_R$ compositeness or both.}
\end{figure}
\clearpage

\end{document}